\documentclass[useAMS,usenatbib]{mn2e}

\usepackage{wasysym}
\usepackage{epsfig}
\usepackage{multirow}
\usepackage{color}

\usepackage{amssymb,amsmath}
\usepackage{appendix}
\definecolor{gold}{rgb}{1.,0.647059,0.}
 \definecolor{purple}{rgb}{0.627451,0.12549,0.941176}
 \definecolor{darkgreen}{rgb}{0.133333,0.545098,0.133333}
\definecolor{marroncaca}{rgb}{0.6,0.2,0}
\definecolor{orangee}{rgb}{1.,0.6,0.4}
\definecolor{olivegreen}{rgb}{0.419608,0.556863,0.137255}

\title[Disc radii variation in black-hole binaries]{On the variation of black hole accretion disc radii as a function of state and accretion rate}
\author[C. Cabanac, R. P. Fender, R. J. H. Dunn and E. G. K\"ording]{C. Cabanac$^{1}$\thanks{E-mail:
c.cabanac@astro.soton.ac.uk (CC)}, R. P. Fender$^{1}$, R. J. H. Dunn$^{1, 2}$\thanks{Alexander von Humboldt Fellow} and E. G. K\"ording$^{1, 3}$\\
$^{1}$School of Physics and Astronomy, University of Southampton, Southampton SO17 1BJ, UK\\ $^{2}$ Excellence Cluster Universe, Technische Universit\"at M\"unchen, 85748 Garching, Germany\\ $^{3}$ AIM - Unit\'e Mixte de Recherche CEA - CNRS - Universit\'e Paris VII - UMR 7158, CEA-Saclay, Service d'Astrophysique,\\ F-91191 Gif-sur-Yvette Cedex, France 
}
\begin{document}

\date{Accepted 2009 April 02. Received 2009 April 02; in original form 2008 November 27}

\pagerange{1-30} \pubyear{2009}

\maketitle

\label{firstpage}

\begin{abstract}
In response to major
changes in the mass accretion rate within the inner accretion flow, Black hole binary transients undergo dramatic evolution in their X-ray
timing and spectral behaviour during outbursts.  In
recent years a paradigm has arisen in which `soft' X-ray states are
associated with an inner disc radius at, or very close to, the
innermost stable circular orbit (ISCO) around the black hole, while in
`hard' X-ray states the inner edge of the disc is further from the
black hole. Models of advective flows suggest that as the X-ray
luminosity drops in hard states, the inner disc progressively recedes,
from a few gravitational radii ($\rm R_{\rm g}$) at the ISCO, to hundreds
of $\rm R_{\rm g}$.  Recent observations which show broad iron line
detections and estimates of the disc component strength suggest that a
non-recessed disc could still be present in bright hard states. In
this study we present a comprehensive analysis of the spectral
components associated with the inner disc, utilising bright states
data from X-ray missions with sensitive low-energy responses(e.g. {\em
Swift, SAX}), including reanalyses of previously published results.  A
key component of the study is to fully estimate systematic
uncertainties associated with such spectral fits.  In particular we
investigate in detail the effect on the measured disc flux and radius of having a hydrogen column
density that is fixed or free to vary.  We conclude that at
X-ray luminosities above $\sim 0.01$ of the Eddington limit, systematic
uncertainties only allow us to constrain the disc to be $\la 10R_{\rm
g}$ from spectral fits.  There is, however, clear evidence that at X-ray
luminosities between $10^{-2}$--$10^{-3}$ of the Eddington rate, the
disc does begin to recede. We include measurements of disc
radii in two quiescent black hole binaries at bolometric luminosities
of $<10^{-7}$ Eddington, and present the inferred evolution of disc
luminosity, temperature, inner radius and accretion rate/efficiency
across the entire range of bolometric luminosities $10^{-8}$--$1$
Eddington. We compare our results with theoretical models, and note
that the implied rate of disc recession with luminosity is consistent
with recent empirical results on the X-ray timing behaviour of black
holes of all masses.
\end{abstract}


\section{Introduction}
The large changes in the X-ray spectra, radio fluxes and timing
 properties observed among the states of the BHBs are commonly
 explained with models involving a variation of the accretion efficiency
 and rate.
This variation of the efficiency can be caused by the inward or outward motion of the inner
radius of the optically thick accretion disc $R_{\rm in}$ (see e.g.,
\citealp{esin97}). Recently, a number of papers have been published
contesting this radius evolution.

Since the first X-ray spectral studies of Cyg X-1
\citep{tananbaum72} and A0620-00 \citep{coe76}, BHBs in outburst seem
to transit between what are now commonly identified as ``canonical
states''.  Those variations were subsequently associated with changes
in radio emission and the timing characteristics of the X-ray
emission.  A dichotomy could clearly be identified, resulting in the
the characterisation of two states.  In one state the spectrum is dominated by a thermal optically
 thick component, is hereafter called the soft or ``Thermal Dominant" (TD)
 \citep{mcclintock03b} state.  The
 other state, the so-called \textit{hard} state, has a powerlaw shape X-ray
 spectrum extending up to a few hundreds of keV (see e.g.,
 \citealp{mcconnell02} for the differences observed in Cyg X-1
 spectra).

Those objects exhibit ejections that are observed in radio and
infrared and that occur in particular states \citep{fender04}.
Although steady and compact jets are typical of the hard state, the
ejection process appears to be quenched whenever it enters in the soft
state \citep{fender99}. \cite{fenderbg04} give simple physical
interpretations of the observed correlations, resulting from which
promising elaborate models based on accretion-ejection solutions in
magnetised discs have been built-up (see e.g., \citealp{ferreira06}
and references therein for an auto-similar analysis of the problem,
but also \citealp{machida06}).

For the vast majority of models invoked, the transition between the
two main canonical states has been interpreted as follows.  In the
soft state, usually occurring at a high luminosity (i.e. $0.01<L_{\rm
  bol}/L_{\rm Edd}<1$), the geometrically thin and optically thick
accretion disc is believed to reach the Innermost Stable Circular
Orbit\footnote{The ISCO isequal to $6\ \rm R_{\rm g}$ for a
  Schwarzschild BH, or $2\ \rm R_{\rm g}$  for a fully spinning Kerr
  BH.} (ISCO). However, in order to reproduce the decreasing disc
efficiency observed in the hard states, the inner part of the disc is
truncated and replaced by a radiatively inefficient and optically thin
flow.  The accreted mass is either advected towards the central
black-hole (ADAF and equivalent: \citealp{esin97, narayan97}) or part
of it is ejected in outflows, for example, an ADIOS
\citep{blandford99} or a Jet Emitting Disc (JED, \citealp{ferreira06})
are examples of structures that could explain the observed behaviour.

XMM-Newton observations of GX 339-4 \citep[hereafter M06]{mi06a} cast
doubt on this accepted interpretation by showing that a broad iron
line together with a dim, hot thermal component was present in its
spectra during the hard state. This effect seems to be observed in a
few other sources such as Cygnus X-1, and SWIFT J1753.5-0127 (M06,
\citealp{mi06b}). However, all of these conclusions are based on
single observation without studying the overall evolution during an
outburst of the source.

A robust way of evaluating the disc inner geometry value is to choose
the same approach as \citet{gd04}. They followed the disc parameter
value during the outburst rise and decline. In the simplistic
hypothesis of an optically thick disc extending down to the last
stable orbit of the black-hole, one would expect that the relationship
between the disc luminosity $L_{\rm disc}$ and the inner temperature
$T_{\rm in}$ should be monotonic; $L_{\rm disc}=K_{\rm disc}T_{\rm
  in}^4$ with $K_{\rm disc}$ constant, whatever the accretion rate.
When the value of $R_{\rm in}$ is much lower than $R_{\rm out}$, then
$K_{\rm disc}\sim 4\pi R_{\rm in}^2\sigma_B$.

Recently, \cite{ry07} (hereafter R07) performed a similar analysis on
the black hole candidate XTE J1817-330.  They reported that no
noticeable changes could be observed in the inner disc radii values,
even when reaching the hard state. On the contrary, taking into
account the effect of irradiation, \cite{gd08} argue that a
significant increase in $R_{\rm in}$ could be measured.  One of the
goals of this paper is to disentangle this apparent discrepancy.

\subsubsection*{Source sample and data selection criteria.}
Considering the large amount of data available in archives, it was
necessary to select the data used in this study.  The data were
selected according to the following criteria: first, we focused on
X-ray transients, as they allow to follow a source on broader
luminosity range. As we focus on the lowest luminosity states and the study of the thermal component, data coming from
instruments with good sensitivity at low energy (typically under 1 keV)
were preferred, ruling out e.g. RXTE archives. For these reasons, the
Swift/XRT data seemed to be ideal as it tends to follow the outburst
of a source by taking a number of different snapshots over time. A
major part of this analysis will therefore be based on data from this
instrument. Lastly, confirmed black hole binaries were favoured, as we
wanted to make comparisons among different sources using
e.g. Eddington luminosities and hence masses and distances.  However,
as the outbursts of XTE J1817-330 and Swift J1753.5-0127 were well
sampled, we also added them into our sample even though they have not
been confirmed as black hole binaries. According to these criteria, our sample reduces to 6 sources:  {\bf XTE J1118+480, GX 339-4, GRO J1655-40, XTE J1817-330, Swift J1753.5-0127 and A 0620-00} (see Tab. \ref{tab-mass-dist-angle} for the adopted masses, angles and distances).

\begin{table}
\centering 
 \begin{minipage}{0.5\textwidth}
\renewcommand{\thefootnote}{\alph{footnote}}
\caption{Orbital parameters used for the whole study.}
\begin{tabular}{@{}lccr@{}}
\hline
 Source&Mass ($\mathrm{M_{\odot}}$)&Distance (kpc)&i ($\rm ^o$)\\
\hline
J1118+480&($8.53\pm0.6$)\footnotemark[1]&$(1.72\pm0.10$)\footnotemark[1]&($68\pm2$)\footnotemark[1]\\
GX 339-4&($7^{12}_{5.8}$)\footnotemark[2]&($8^{9.4}_{6.7}$)\footnotemark[3]&($30^{60}_{20}$)\footnotemark[4]\\
J1655-40&($6.3\pm0.5$)\footnotemark[5]&($3.2^{3.4}_{1.7}$)\footnotemark[6]&($70^{71}_{64}$)\footnotemark[7]\\
J1817-330&($6^{10}_{3.5}$)\footnotemark[8]&($6.3^{10}_{2.6}$)\footnotemark[8]&(60)\footnotemark[8]\\
J1753.5-0127&($6^{12}_5$)\footnotemark[9]&($2.9^{10}_{2.9})$\footnotemark[9]&(60)\footnotemark[9]\\
A0620-00&($11\pm1.9$)\footnotemark[10]&($1.16\pm0.11$)\footnotemark[10]&($40.75\pm3$)\footnotemark[10]\\
\hline
\label{tab-mass-dist-angle}
\end{tabular}
\footnotetext[1] {\cite{gelino06}.}						
\footnotetext[2] {\cite{hynes03} for the mass function.}
\footnotetext[3]{\cite{zdziarski04}.}	
\footnotetext[4]{Compatible with \cite{cowley02}.}				
\footnotetext[5]{\cite{green01}}						
\footnotetext[6]{\cite{hjellming95} for the value but we used \cite{foellmi06} as a lower limit.}					
\footnotetext[7]{\cite{vanderhooft98}.}	
\footnotetext[8]{S07 for the mass and distance, however see the text for the lower limit in distance. Arbitrary inclination.}	
\footnotetext[9]{Mostly arbitrary but average values expected for BHB. See \cite{mi06b} for the low limit in distance.}
\footnotetext[10]{\cite{gelino01}.}
\end{minipage}

\end{table}
\renewcommand{\thefootnote}{\arabic{footnote}}
\section[]{Data reduction and reanalysis}\label{sec-data-red}

The purpose of our paper is to extend previous studies of the disc
parameters to
lower disc luminosities, as well as to systematically and uniformly test
spectral models.  There are two approaches that can be chosen to overcome the difficulties of
determining the parameter values of the disc. The first is to try to
fit the spectra with the best models available. This means using
sophisticated comptonisation codes or tables to fit the hard
component, and the possible reprocessing of hard X-rays on the optically
thick disc. The main advantage is to usually obtain better fits to the
spectra.  This comes at a price, of having a larger number of free parameters. The other main
drawback when adopting this approach is that comparison among
sources or different observations is difficult as the models chosen
are not necessarily the same.

The other method consists of using the simplest model, involving the
least number of parameters.  This allows the same simple model to be
fitted to the spectra resulting from every source and observation. The price which is
paid is a lower goodness of fit, but it allows comparisons among
sources to be made.

We choose to adopt this latter approach in our study.  The
spectra were fitted with a powerlaw at high energy and any possible
soft component by a multicolour disc. We also added the photoelectric
absorption of the neutral hydrogen in the line of sight. Therefore the number of
free parameters in the model does not exceed five: the absorption
$N_{\rm H}$, the multicolour disc temperature $T_{\rm in}$ and
normalisation $K_{\rm disc}$, and the powerlaw photon index $\Gamma$
and normalisation $K_{\rm PL}$. However, in order to
  study the effect of changing the high energy emission model on
  the conclusions drawn, we also used a thermal comptonisation model
  (\texttt{compTT} in {\sc xspec}) for some of the observations. As the value of the
  high-energy cutoff was usually unavailable, we therefore fixed the
  value of the temperature of the thermal electron to 50 keV unless
  stated otherwise. The number of free parameters are therefore still
  reduced to two for this component: the optical depth and the
  normalisation.

\subsection{Precision of the cross-analysis}

In this paper we pay particular attention to comparisons
among several sources (Section \ref{sec-multi}), which involves
different orbital parameters and distances, disc radius or
luminosities have all been rescaled in terms of the gravitational
radius $\rm R_{\rm g}=GM/c^2=1.48M/\mathrm{M_{\odot}\ km}$ and
Eddington luminosity $L_{\rm Edd}=1.48 \times
10^{38}(M/\mathrm{M_{\odot}})\ \mathrm{erg\ s^{-1}}$ (e.g.,
\citealp{gd04}). Depending on the disc inclination and for projection
reasons we use: 

\begin{align}
\frac{L_{\rm disc}}{L_{\rm Edd}} &= \frac{F_{\rm disc}}{L_{\rm Edd}}\times\frac{2\pi d^2}{\cos(i)}\\
&= 4\times  10^5\times  \frac{F_{\rm disc}\,{d_{\rm kpc}}^2}{M_{\rm \mathrm{M_{\odot}}}\cos(i)},
\end{align}     
where $d_{\rm kpc}$ is the distance expressed in kpc, $F_{\rm disc}$
the disc luminosity in $\mathrm{erg\ cm^{-2}\ s^{-1}}$, $M_{\rm
  M_{\odot}}$ the black hole mass expressed in solar masses, and $i$
the inclination angle ($i=0$ is a ''face-on'' disc). 

Assuming that the hard component emits isotropically, we have approximately:
\begin{equation}
\frac{L_{\rm hard}}{L_{\rm Edd}} = 8\times  10^5\times  \frac{F_{\rm hard}\,d_{\rm kpc}^2}{M_{\rm \mathrm{M_{\odot}}}},
\end{equation}
For the evaluation of the disc parameters, and for the reasons
explained above, we use the simple multicolour disc model
(\texttt{diskbb} in {\sc xspec}, see e.g.,
\citealp{mitsuda84}). Several drawbacks have already been pointed out
concerning this model as it neglects processes such as increased
scattering at high temperatures, or the differential emission that can
occur between different altitudes in the disc (the central part of the
disc may be warmer). Such drawbacks mainly affect the effective
temperature compared to the observed one. However, the temperature
shift is only 4\% for a 0.5 keV disc and this effect decreases with
decreasing temperature \citep{gd04}. The fact that we aimed to probe
the lowest luminous states also motivated our choice to use a simple
\texttt{diskbb} to model the thermal component. 

As $R_{\rm in}$ is obtained via the normalisation ($K_{\rm
  disc}=(R_{\rm in,\ km}/d_{\rm 10\ kpc})^2\cos(i)$ in {\sc xspec}) of
the multicolour disc model, it gives:

\begin{equation}
\frac{R_{\rm in}}{\rm R_{\rm g}}=0.677\left(\frac{d}{10\ \mathrm{kpc}}\right)\left(\frac{M}{\mathrm{M_{\odot}}}\right)^{-1}\left(\frac{K_{\rm disc}}{\cos(i)}\right)^{0.5}.
\end{equation} 
Unless mentioned explicitly in the text, all errors in this paper will
also be expressed in terms of 90\% (1.64 $\sigma$) confidence range
(including the plots).

\subsection{XTE J1118+480: multiwavelength campaigns}

\subsubsection{In outburst}

For the study of the BHB XTE J1118+480, we mainly used the parameters
given in the literature. The number of observations where a thermal
component is required in the spectra are quite scarce, as there is
just the one coming from the 2000 outburst and one other, during
quiescence in 2002. Among the articles dealing with this outburst
where a thermal component is observed, two were based on a
multi-wavelength study including UV data (\citealp{mcclintock01} and
\cite{chaty03}, hereafter (MC01) and (Ch03)) whereas the other focused
on results given by Beppo-SAX data \citep{frontera03}.

We note a particular issue regarding the parameter of the disc
obtained by MC01. There seems to be an inconsistency between the low
value of the radius given in the text when using a simple multicolour
disc to fit the HST and EUVE data, and the spectrum given in their
Fig. 3.  They obtain an internal radius of $34\ \rm R_{\rm S}$ ($\rm 1 R_{\rm S}=2 R_{\rm g}$), and an
internal temperature of 24 eV. Such a model, (assuming as (MC01) do,
$M=6\mathrm{M_{\odot}}$, $d=1.8\ {\rm kpc}$ and $i=80^{\rm o}$) should
peak at a value of $\nu F_{\rm \nu}\sim 10^{-11.8}\  {\rm
  erg\ cm^{-2}\ s^{-1}}$. From their Spectral Energy Distribution (SED)  it is clear that this is
about two orders of magnitude lower than the peak value shown in the
third panel (i.e. $\nu F_{\rm \nu}\sim 10^{-8.6}\ {\rm
  erg\ cm^{-2}\ s^{-1}}$). In order to obtain a similar value with
their orbital parameters and temperature, we have to set $K_{\rm
  disc}\sim 7\times  10^8$. This then gives us $R_{\rm in}\sim 300
R_{\rm S}$ when we take into account the orbital values displayed in
our Tab. \ref{tab-mass-dist-angle}. 

This is moreover consistent with the value of $R_{\rm in}\sim 352
R_{\rm S}$ obtained by (Ch03) based on the same data set (HST, EUVE,
CXC, RXTE, UKIRT and Ryle telescope), where SAX and the VLA data were
added, and modelled by the same simple multicolour disc +
powerlaw(s). The small residual differences could come from the value
of the absorption adopted by each authors ($N_{\rm H}=1.1\times
10^{20}\ {\rm cm^{-2}}$ for (Ch03) and $N_{\rm H}=1.3\times
10^{20}\ {\rm cm^{-2}}$ for (MC01)), and the secondary contribution in
optical as well.

However, when using SAX data in their multi-wavelength analysis,
(Ch03) did not take into account the lowest energy bins (0.13-26 keV)
as the flux value conflicts with the EUVE ones. They argued that this
discrepancy could come from calibration issues in SAX responses at low
energy. However, as the amount of available observations for XTE
J1118+480 outburst is quite low, we have tried to include these data in our
study. This is also supported by the fact that in studies based on the
full SAX broad band spectrum \citep{frontera01,frontera03}, a thermal
component was also detected though peaking at a higher energy
($\sim$50 keV) than those given in (MC01) and (Ch03). However in
\cite{frontera03}, the low energy part of the spectrum was fitted with
a single black-body spectrum and thus we tried to probe the effect of
reanalysing the data with a multicolour disc.  

Therefore, we used and reanalysed the processed LECS (low-energy
concentrator spectrometer, 0.1-4 keV, \citealp{parmar97}), MECS
(medium-energy concentrator spectrometer, 1.8-10.5 keV
\citealp{boella97}) and PDS (Phoswich Detector System, 10-200 keV,
\citealp{frontera97}) spectra available on the archive for both
``ToO1'' (ObsId 21173001) and ``ToO3'' (ObsId 211730012) observations
mentioned in \cite{frontera03}. The usual response files were used, we
allowed a free normalisation among the three instruments and one
per cent systematics were also added. As pointed out by
\cite{frontera03}, the addition of a thermal component to the absorbed
cutoff powerlaw model is necessary. When we add a \texttt{diskbb} (the
absorption being free to vary), the reduced $\chi^2$ drops from
$\chi^2/\nu=286.05/191=1.50$ to 213.64/189=1.130 (F-test probability
$=\ 1.05\times  10^{-12}$) in ToO1 (from $\chi^2/\nu=2.98$ to 1.25 in
ToO3) . We noted also that taking a single black body for the thermal
component gives similar goodness of fit as in that case
$\chi^2/\nu=212.60/189=1.126$ in ToO1 for example. As stated before,
we thus kept the results coming from a multicolour disc. The 
results of the fits to the disc parameters are given in
Tab. \ref{tab-resultsc} of the appendix.

We note that the results we obtain for the thermal component are quite
different between the SAX data and the EUVE data, as was pointed out in
(Ch03). The value of the disc flux is about 10 times lower in SAX
data, the temperature is about three times higher than in the
EUVE. However the major difference comes from the normalisation of the
multicolour disc as it is $10^3$ lower. This means the disc is closer
to the central black hole by a factor 30. As there is no obvious
reasons to favour one or the other set of data, we decided to keep
both of them, and infer that the real state of the disc should be
between both. In the following figures, results coming from both data
sets are displayed explicitly. 

\subsubsection{In quiescence}

For the analysis in quiescence, we mainly based our analysis
on the results coming from \cite{mcclintock03a} (MC03a). Compared to
those authors, we did some slight corrections to estimate disc radii
and luminosities for XTE J1118+480, given the accuracy of the orbital
parameters obtained since then by \cite{gelino06}. For example, with
the value of the internal radius that (MC03a) obtain (around 3100 $\rm
R_{\rm g}$ but for $M=7\mathrm{M_{\odot}}$, $d=1.8\rm\ kpc$ and
$i=80^o$), we derived the corresponding \texttt{diskbb} normalisation
($K_{\rm disc}$).  

We then added the temperature estimate to evaluate the net disc flux
contribution ($F_{\rm disc}\propto K_{\rm disc}T^4$). Subsequently we
corrected the luminosities and the internal radius from the newest
orbital parameter in order to obtain $R_{\rm in}\sim 1640 \rm\ R_{\rm
  g}$, $L_{\rm hard}\sim 1.3\times 10^{-8}\ L_{\rm Edd}$ and $L_{\rm
  bol}\sim 7.7\times 10^{-8}\ L_{\rm Edd}$.  Note that, except for the
disc temperature, no statistical errors coming from the fits were
available for the luminosities nor the radius inferred and hence those
latter are not displayed in the figures of the study.  
 
\subsection{A0620-00: HST/STIS and Chandra}

In order to estimate the disc properties of A0620-00, we used the
values published by (MC03a), \cite{mcclintock00} and
\cite{mcclintock95} as following.  We have already pointed out that the
lack of optical-UV observation of the source (compared to XTE
J1118+480) is still a major problem as we are not sure that a
multicolour disc is the best model to fit the data. We noted
that MC03a, \cite{mcclintock00} interpreted those data in the
framework of an ADAF, whereas in our study we try to explain the
possible UV extra component in the context of a multicolour disc.

Another limitation concerning this source comes from the fact that the
X-ray and optical observations were not simultaneous. Therefore the
conclusions resulting from this object must be taken with caution. 

However, concerning the disc properties, and due to the simplicity of
the model, we only had to find a rough estimate of the inner
temperature and the normalisation of the \texttt{diskbb}. For the
temperature, we note that the HST/STIS disc spectrum obtained by
\cite{mcclintock00} in 1998 was very similar to the one obtained six
years before (comparing Fig. 2 of \citealp{mcclintock00} and Fig. 5 of
\citealp{mcclintock95}). Hence, we took the 9000 K ($7.7\times
10^{-4}\ \rm keV$) black body spectrum \citep{mcclintock95} as a
reference for the temperature value. A \texttt{diskbb} spectrum with
such a temperature value peaks nearby 3500 \r{A}, as it was also
observed in the 1998 spectrum.

This temperature value fixed, we then obtained the normalisation of
the \texttt{diskbb} model via the flux value at $4\times 10^{-3}\ \rm
keV$, which is around $\nu F_{\rm \nu}\sim
10^{-12.4}\ \mathrm{erg\ cm^{-2}\ s^{-1}}$. It thus gave us $K_{\rm
  disc}\sim 2.8\times 10^{11}$. Hence, the corresponding radius would
be (following mass, distance and angle value of
Tab. \ref{tab-mass-dist-angle}): $R_{\rm in}=4.5\times 10^3\ \rm
R_{\rm g}$. The bolometric flux of the disc component is $F_{\rm
  disc}=2.3\times 10^{-12}\ \mathrm{erg\ cm^{-2}\ s^{-1}}$.

For the hard X-ray component, we took the powerlaw model given by
(MC03a) where the photon index value is given as $\Gamma=2.26$. We
inferred the normalisation value of the powerlaw considering that the
flux value at 1 keV is  $\nu F_{\rm \nu}\sim
10^{-14}\ \mathrm{erg\ cm^{-2}\ s^{-1}}$ (see Fig. 11 of MC03a). The
corresponding absorbed ($N_{\rm H}=1.94\times 10^{20}\ {\rm cm^{-2}}$,
see MC03a) bolometric (0.02-200 keV) flux is then $F_{\rm
  pl}=4.3\times 10^{-14}\ \mathrm{erg\ cm^{-2}\ s^{-1}}$.

\subsection{GX 339-4: ASCA/GIS data}
For the three ASCA/GIS observation of GX 339-4 (see \citealp{wilms99}
for a previous study), we also explored the influence of the $N_{\rm
  H}$ value on the disc parameters. Again, for consistency we fitted
the data by a simple absorbed \texttt{diskbb+powerlaw}
(\citealp{wilms99} used a broken powerlaw, a choice that might influence
the radius estimate). We chose to probe this effect using three
different ways.  Firstly by fixing the absorption value to the one
adopted by \cite{wilms99}, i.e. $N_{\rm H,\ 1}=0.62\times
10^{22}\ {\rm cm^{-2}}$ (see also \citealp{zdziarski98}), then by
fixing it to the weighted mean average value we obtained in the
following study with \textit{Swift} data (see
Section \ref{subsec-influ-nh}), i.e. $N_{\rm H,\ 2}=0.43 \times 10^{22}
{\rm cm^{-2}}$, and finally by leaving it free to vary. Results of
those fits are given in Tab. \ref{tab-asca-339}. 

\begin{table*}
\centering 
\caption{Results of the fits by an absorbed \texttt{diskbb+powerlaw} of GX 339-4 ASCA/GIS spectra when changing the absorption value.}
\begin{tabular}{|c|c|c|c|c|c|c|}
\hline
&$N_{\rm H}$ value &$K_{\rm disc}$&$kT_{\rm in}$&$F_{\rm disc}$&$F_{\rm pl}$&$\chi^2/\nu$\\
\hline
\multirow{3}*{Obs \#1}&0.62&$8.2^{16}_{4.6}\times 10^3$&$0.19\pm0.02$&$2.3^{2.7}_{1.5}\times 10^{-10}$&$9.0\pm0.3\times 10^{-10}$&1054.03/954\\
                     &0.43&$1.1^{6.1}_{0.29}\times 10^3$&$0.21^{0.27}_{0.17}$&$5.1^{6.3}_{0}\times 10^{-11}$&$1.003^{1.006}_{0.997}\times 10^{-9}$&1030.70/954\\
                     &$0.32\pm0.02$ (free)&\multicolumn{4}{c|}{powerlaw only sufficient}&1025.03/954\\
\hline
\multirow{3}*{Obs \#2}&0.62&$2.4^{4.0}_{1.6}\times 10^{4}$&$0.18\pm0.01$&$5.5^{6.0}_{4.7}\times 10^{-10}$&$1.55\pm 0.05\times 10^{-9}$&1279.46/1174\\
                     &0.43&$7.8^{40}_{1.9}\times 10^{3}$&$0.17\pm0.04$&$1.4^{1.4}_{0.}\times 10^{-10}$&$1.75^{1.81}_{1.70}\times 10^{-9}$&1201.70/1174\\
                     &$0.29\pm0.01$ (free)&\multicolumn{4}{c|}{powerlaw only sufficient}&1184.69/1175\\
\hline
\multirow{3}*{Obs \#3}&0.62&$3.37^{3.89}_{3.04}\times 10^4$&$0.211^{0.214}_{0.206}$&$1.38^{1.49}_{1.37}\times 10^{-9}$&$4.00^{4.04}_{3.93}\times 10^{-9}$&2970.47/1469\\
                     &0.43&$2.68^{3.16}_{2.30}\times 10^{3}$&$0.279^{0.288}_{0.270}$&$3.42^{3.63}_{3.41}\times 10^{-10}$&$4.63^{4.71}_{4.55}\times 10^{-9}$&2506.09/1469\\
                     &$0.27\pm0.01$ (free)&$135^{190}_{100}$&$0.46^{0.50}_{0.26}$&$1.31^{1.46}_{1.15}\times 10^{-10}$&$5.37^{5.53}_{5.24}\times 10^{-9}$&2366.47/1468\\
\hline
\label{tab-asca-339}
\end{tabular}
\end{table*}

As noticed with the \textit{Swift} data, the value of the disc
normalisation tends to decrease whenever the absorption does. For
example in the third observation the corresponding internal radius
would reach a value of $\sim17\ \rm R_{\rm g}$ when the value of the
absorption is fixed to $N_{\rm H}=0.62\times 10^{22}\ {\rm cm^{-2}}$,
and drops to $\sim 1\ \rm R_{\rm g}$ when it is left free to vary. The
disc is not even necessary in order to fit the spectra properly when
using a floating $N_{\rm H}$ in observation one and two. However, it is
unclear whether this is due to a real trend or a systematical effect
linked to ASCA/GIS data analysis.

A good way of disentangling this problem would consist in applying the
same processes we did for \textit{Swift} data (see
Section \ref{subsec-influ-nh}). However, as the the number of
observations is quite low in this case (3 observations compared to
more than 20), we would not be able to draw definite conclusions. As a
result of this, we chose to keep the results coming from the case when
$N_{\rm H}$ is fixed to $0.43\times 10^{22}\ {\rm cm^{-2}}$ (average
value obtained with \textit{Swift}) for the following study.

\subsection{XTE J1817-330, SWIFT J1753.5-0127, GRO J1655-40 and GX
  339-4: \textit{Swift} data}

For XTE J1817-330, we processed the data in the same way as in R07
(especially concerning the region sizes for the pile-up correction)
except that we used the XRT pipeline software version 0.11.5
(2007-08-23 release date) and more recent response files coming from
the CALDB 20071101 version (v009) as well. Of particular importance
are the better corrections to the residuals under 0.6 keV in WT mode
compared to earlier response files. It allows us to perform spectral
fitting within the 0.3-10 keV energy range though adding 3\% of
systematics \citep{campana07}.

For GX 339-4, GRO J1655-40 and SWIFT J1753.5-0127 especially in WT
mode, the pile-up correction was done using the second method as
described in \cite{mineo07} and recapped here. For a given observation, different annuli
(PC mode) or boxes (WT mode) of decreasing sizes were extracted in the
images and spectra were generated and fitted with the most convenient
model (either a powerlaw or a powerlaw+disc). Then when the fit
parameter reach the asymptotic values obtained with the smallest
annuli/boxes\footnote{In our case, if the photon index differences are
  less than 5 per cent of the asymptotic value.}, it determines the maximal
size of the admitted region.

For spectra obtained in LrPD mode (GRO J1655-40, see also
\citealp{bro06}), spectra have been fitted in the 0.5-10 keV
range. However, we noticed that the value of 5 per cent given
  by \cite{cusumano05} tends to lead to particularly low $\chi^2$
  values.  We obtain an average value of the reduced $\chi^2$ of
  about $\sim 0.6$ on 13 observations.  We therefore investigated the
  effect of changing the value of the systematics on the
  calculation of the errors on the fit parameters.  Changing the
  systematics from 2.5 per cent
  to 5 per cent,
  the relative errors on e.g. the absorption value in obs. 00030009005
  jumps from only 1.1 per cent to 1.7 per cent. The disc normalisation
  changes from 3.1 per cent to 4.6 per cent. Therefore, if the systematics seem
  overestimated in LrPD mode, the actual value has little effect
  on our study.  We therefore kept the 5 per cent given in \cite{cusumano05}.

 For certain observations where the disk is dominating, the addition
 of a powerlaw gives an unrealistic photon index value (sometimes
 negative). For those latter cases, we did not take into account the
 7.5-10 keV energy range (as \citealt{bro06} do). The addition of a
 powerlaw is thus sometimes not required. Spectra were also binned in
 order to get at least 20 counts/channel. 

For any given observation of the source, there can exist gaps due to
Good-Time-Interval (GTI). In the early version of the XRT data
processing software somewhere the spectra for each GTI had to be
extracted separately, this effect is now taken into account. However,
we decided to keep this process when fitting spectra, i.e. for one
observation we can get several spectra and thus different fit
parameter values. As we wanted to probe any effect on the disc even
when quite dim, this way of processing allows to check if any change
in the disc parameters would be due to a temporary instrumental bias.

\subsection{Fits and flux estimate methods}

Unless explicitly mentioned, all spectra in this study were fitted
using {\sc xspec} v12.3.1ao \citep{arnaud96}. We also used the
\texttt{fluxerror} tcl script based on Monte-Carlo method that was
provided by K. Arnaud.  It allows the estimation of flux errors on each
component of the model separately. However, we slightly modified the
script in order to be able to compute the errors within an extended
energy range instead of the default one. In order to estimate the
bolometric luminosity, absorbed flux computation has thus been
performed between 0.05 and 200 keV for the hard component. This choice
was motivated by the fact that:

\begin{enumerate}

 \item the cut-off observed in hard states generally occurs around a
   few hundred keV or the flux contribution over 200 keV is low for
   typical soft state (with $\Gamma\sim 2.5$, $F_{\rm
     pl,\ 200.-20000\ keV}/F_{\rm pl,\ 2.-200\ keV}=0.1$). 

\item  there should also exist a ``cut-off'' of the powerlaw at low
  energy (taken into account when fitting by e.g., thermal
  comptonisation models) that can be roughly mimicked by the
  absorption. 

\end{enumerate}
A summary of the fit results used in this study for the disc geometry
analysis (i.e. Sections \ref{subsub-nh-disk}, \ref{sec-multi} and
\ref{sec-discussion}) are displayed in Tabs. \ref{tab-resultsa},
\ref{tab-resultsb} and \ref{tab-resultsc}.

\section{The Hydrogen Column Density: To fix or not to fix?}
\label{subsec-influ-nh}

The value of the hydrogen column density adopted is always a central
problem when trying to constrain the properties of the disc.  For many
previous studies (R07, \citealp{gd04}) this value was fixed, either by
referring to the most common adopted values  (e.g., \citealp{gd04}),
or by taking the one inferred by the fit at the highest fluxes.

Studying the geometrical properties of the optically thick disc via
X-ray analysis is always a challenge as one has to determine the value
of two parameters that appear to be quite correlated - the column density
and the normalisation of the emitting disc (that is proportional to
its projected area). Moreover, this degeneracy will be strengthened by
intrinsic and/or technical effects; the lower the maximum temperature
of the disc, the higher the column density or the lower the spectral
sensitivity of the instrument observing, the higher the
degeneracy.

For example, even for sources like GRO J1655-40
or GRS 1915+105 that show high disc temperatures during their
outburst, several authors fix the hydrogen column density value when
using RXTE observations (\citealt{gd04} for example). This is
justified as the low efficiency of the PCA under $2\ \rm keV$ would
not constrain easily the value of the absorption.  However, when using
observatories as efficient at low energy as 
 Chandra, XMM or \textit{Swift}, there seems to be fewer ``a priori'' 
 underlying reasons to fix the $N_{\rm H}$ for every spectral
 analysis. 

In this section we will study the effect of letting the value of
$N_{\rm H}$ varying versus keeping it fixed, especially during the transition from soft
to hard state.  We apply this to the \textit{Swift}-XRT data of GX
339-4, XTE J1817-330, GRO J1655-40 and SWIFT J1753.5-0127. Initially
we evaluated the value of $N_{\rm H}$ at the highest luminosities. In
that case, its value varies between $4.6\ \times 10^{21}$ to
$2\ \times 10^{21}\ {\rm cm^{-2}}$ from observation number one to 13
in GX 339-4 data, and between  $2\ \times 10^{21}$ to less than
$1\ \times 10^{21}\ {\rm cm^{-2}}$ for XTE J1817-330 for
example. Then,  \textbf{when a multicolour disc is required} according
to our standards (see hereafter in section \ref{subsub-nh-disk}), we
can compute the weighted mean value $N_{\rm H,\ wmean}$.
 It gives $N_{\rm H,\ wmean} = 0.43\pm 0.02\times 10^{22}\ {\rm
   cm^{-2}}$ for GX 339-4, $ 0.12\pm 0.02\times 10^{22}\ {\rm
   cm^{-2}}$ for XTE J1817-330 (in agreement with the value obtained
 by R07), $0.246\pm0.011\ {\rm cm^{-2}}$ for SWIFT J1753.5-0127 and
 $0.73\pm0.014\ {\rm cm^{-2}}$ for GRO J1655-40.

\begin{figure*}
\vbox to 220mm{\vfil 
\includegraphics[width=0.75\textwidth]{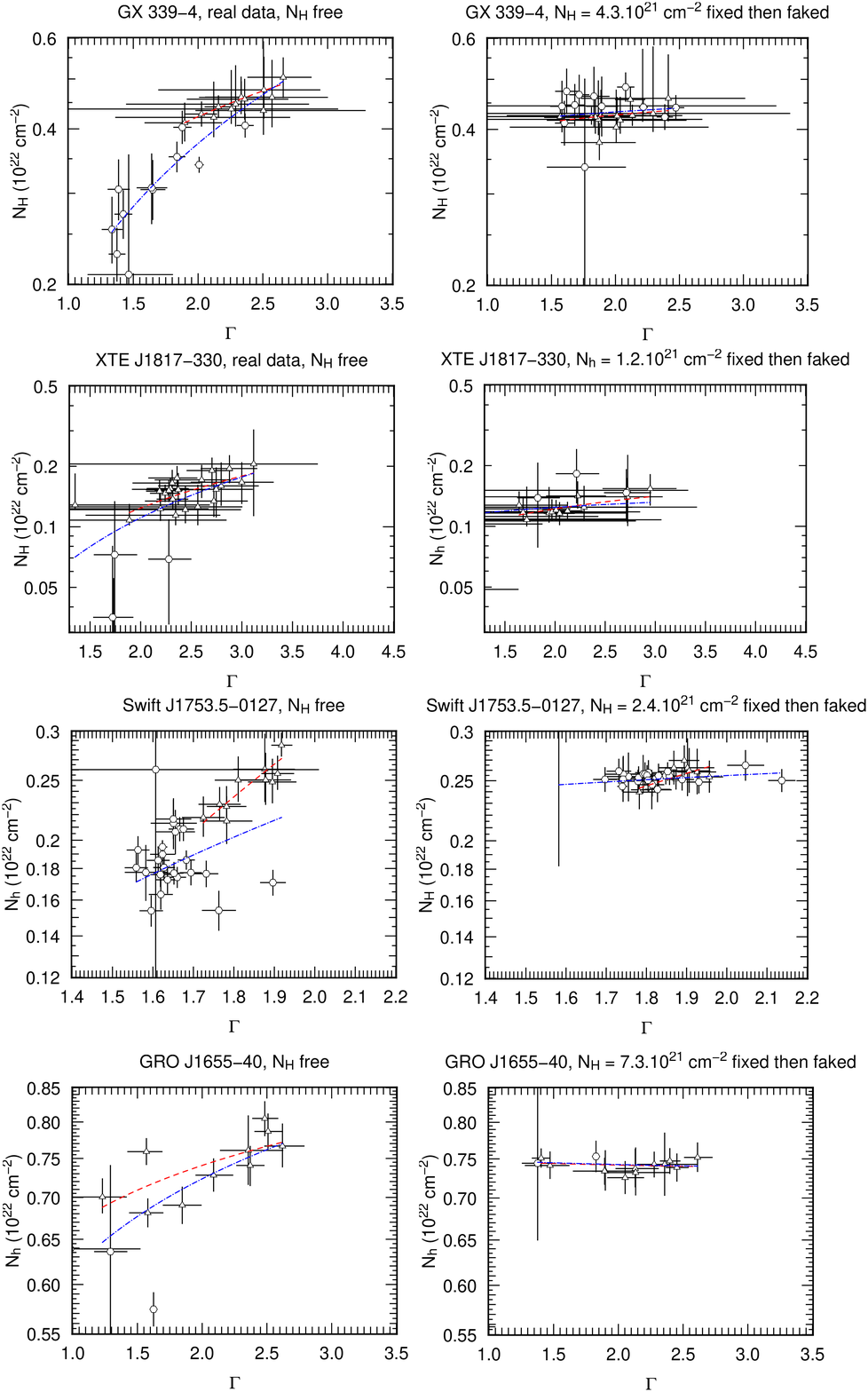}
 \caption{Column density variation during the transition in GX 339-4,
   XTE J1817-330, SWIFT J1753.5-0127 and GRO J1655-40 when high energy
   component in real and faked data are fitted with a powerlaw model.
   The model is either a $wabs*po$ ($\circ$) or a $wabs*(diskbb+po)$
   ($\bigtriangleup$). Dashed line shows the fit by a powerlaw of
   the $\bigtriangleup$ and the dotted-dashed line is a fit of
   all acceptable points ($\bigtriangleup$ and $\circ$ when
   $\Gamma>1$). See Tab. \ref{nh-gamma-tbl} for the fit results. Note
   that for some observations a powerlaw component is not necessary to
   fit the spectra (e.g for GRO J1655-40).  Therefore the number of points
   in those graphs can be different from the numbers given in the
   histograms of Fig. \ref{histo-nh-po}.  
} 
 \label{gamma-nh}
\vfil
} 
\end{figure*}

\begin{figure*}
\vbox to 220mm{\vfil 
\includegraphics[width=0.75\textwidth]{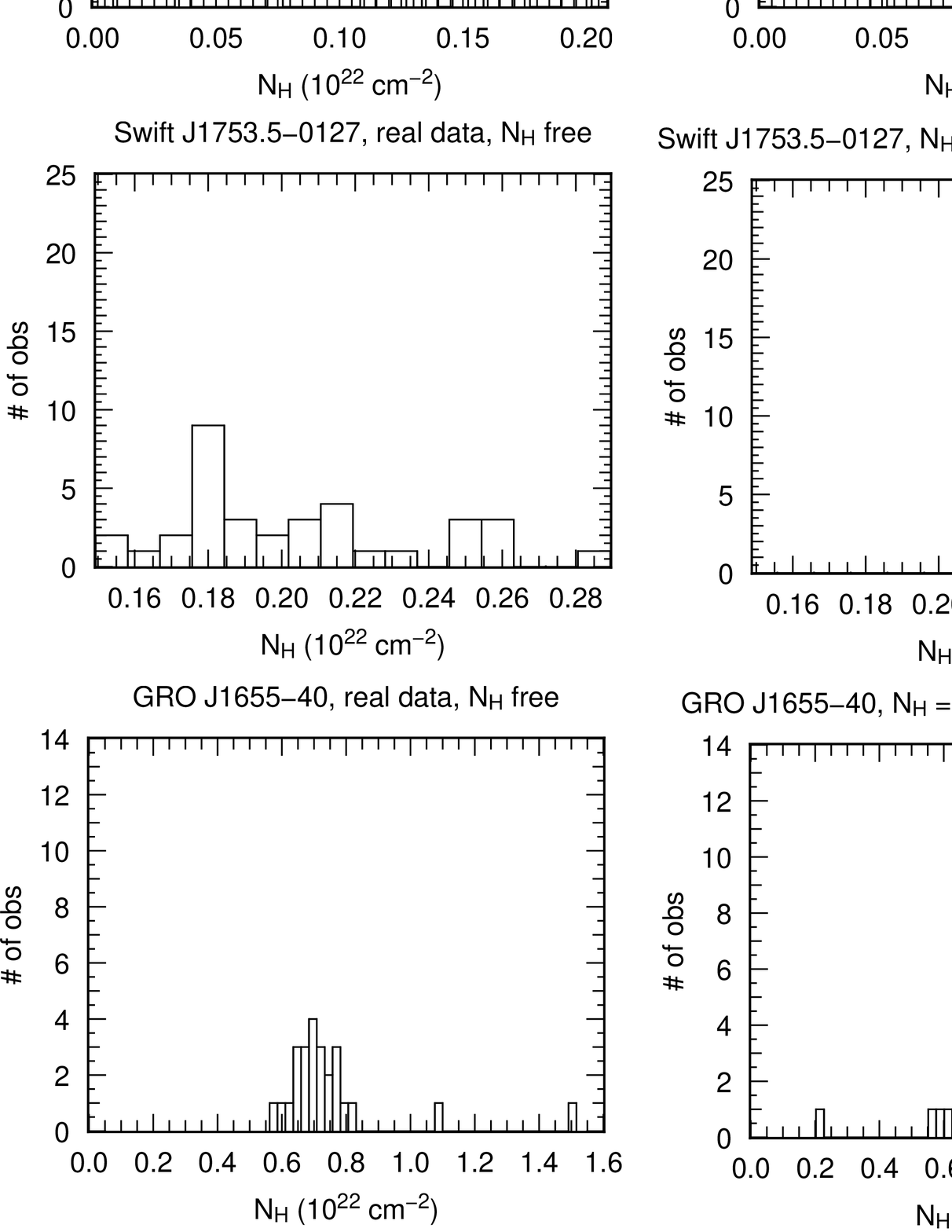}
 \caption{Distribution of $N_{\rm H}$ when the high energy component
   in real and faked data are fitted with a powerlaw model. Left
   panels correspond to the real data fitted with a floating $N_{\rm
     H}$, when the right panels correspond to data obtained by faking
   the real data when $N_{\rm H}$ is fixed to the weighted mean value
   obtained in the highest states. 
} 
 \label{histo-nh-po}
\vfil
} 
\end{figure*}

If we now plot the dependency between the value of $N_{\rm H}$
and the photon index as shown in the first column of
Fig. \ref{gamma-nh}, one can see that there seem to be a slight
correlation between the value of $N_{\rm H}$ obtained from one
observation to another. That effects can also be seen when looking at
the distribution of $N_{\rm H}$ values presented in the left column of
Fig. \ref{histo-nh-po}.

 However, as we combine three different models
 ($\mathtt{wabs}\times(\mathtt{diskbb}\ +\ \mathtt{powerlaw})$ or
 $\mathtt{wabs}\times(\mathtt{diskbb}\ +\ \mathtt{compTT})$) with a
 total of five free parameters, one cannot \textit{a priori} exclude a
 systematic degeneracy among those. Indeed, one can argue that the
 trend between the photon index and the value of $N_{\rm H}$ obtained
 could be purely due to the model used, as a higher photon index would
 require a high absorption to obtain one and the same flux at low
 energies.

\begin{figure*}
\includegraphics[width=1.0\textwidth]{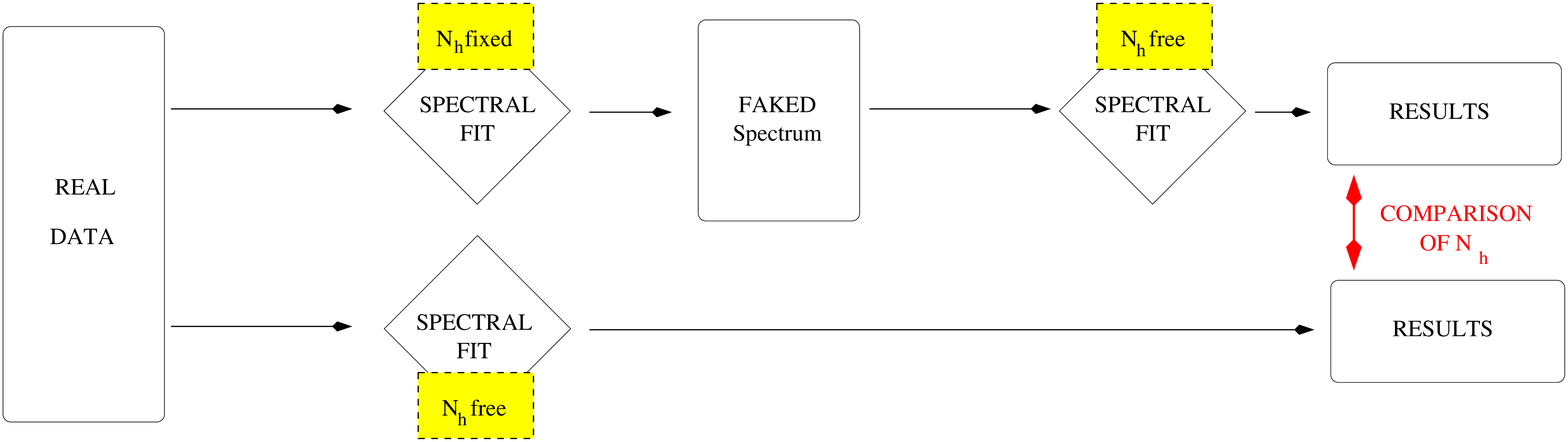}
\caption{Schematic process we adopted to probe if the variation in the column density value observed was a systematic effect of data analysis or real. This comparison is called the \textit{fff method} in the text.} 
 \label{process_faking}
\end{figure*}

Therefore, in order to probe if the trend was real or due to the fitting
process, we proposed to follow the following scheme summarised on
Fig. \ref{process_faking}.  Having processed the \textit{real} data as
described above (i.e. with a 
floating $N_{\rm H}$), in parallel and for each observation we fitted
the same spectrum in the same way (i.e. adding a disc, or not, when it
was not appropriate), \textbf{except that the $N_{\rm H}$ is now fixed
  to the best weighted mean value obtained above}. With those new fit
parameter values, we then faked a spectrum with the {\sc xspec}
command \texttt{fakeit}, taking the same response files and time
elapsed, thus obtaining the so-called \textit{fixed then faked (ff)}
spectrum. In a third step, we fitted this \textit{ff} spectrum with a
floating $N_{\rm H}$, hence obtaining \textit{fff} (\textbf{f}ixed,
\textbf{f}aked, \textbf{f}ree) parameters. The comparisons in the
value of $N_{\rm H}$ obtained in the \textit{fff} and in the simple
case of a floating $N_{\rm H}$ would allow one to determine whether
the observed variation is more consistent with statistic/systematic
deviation or with a real trend. 

Results are displayed from Fig. \ref{histo-nh-po} to
\ref{tau-nh-comptt} and Tabs. \ref{nh-gamma-tbl}, \ref{nh-tau-tbl} and
\ref{nh-ks-test-tbl}. Two complementary approaches have been used:
first by comparing both distributions of $N_{\rm H}$ obtained, and
then the apparent photon index $\Gamma$-$N_{\rm H}$ correlation
observed. As stated in the previous section, both cases where the high
energy part of the spectrum is fitted with either a powerlaw or a
\texttt{comptt} \cite{titarchuk94} model were examined. Both analysis
seems to be useful, as it seems to lead to different conclusions.

\subsection{High energy component fitted by a powerlaw.}

When a powerlaw is used first, one can see in the histograms of
Fig. \ref{histo-nh-po} the relative differences between the
distribution of $N_{\rm H}$ obtained in the real and the faked
spectra. As a result, if we except the case of GRO J1655-40,  the real
distributions seems to be flatter than the \textit{fff} ones. We then
performed a Kolmogorov-Smirnov (KS) test to compare the latter
distributions. The results, shown on Tab. \ref{nh-ks-test-tbl}, give a
low probability that the real and the \textit{fff} spectra could come
from the same distribution. However, except for SWIFT J1753.5-0127,
the KS test probabilities obtained are still too high to draw any
definite conclusions.

Had the fitted $N_{\rm H}$ been an artifact of the degeneracy with the
fit parameter, we would have expected the distribution to be
similar. To confirm the non degeneracy,  we fitted the $N_{\rm H}$
versus $\Gamma$ by a powerlaw in both real and \textit{fff} cases as
shown on Fig. \ref{gamma-nh}. Note that the choice of a powerlaw is
completely arbitrary. However, it seems straightforward to probe
whether there is a systematic trend by examining and comparing the
exponent value obtained: a nil value of this latter would for example
mean that $N_{\rm H}$ is independent of $\Gamma$. We performed this
fit using either the whole dataset (dotted-dashed curve in
Fig. \ref{gamma-nh}), or only the observations where a disc is
required (dashed curve). For the $\chi^2$ evaluation when fitting,
each point in the $N_{\rm H}$ vs $\Gamma$ diagram is weighted
according to its error on $N_{\rm H}$ only. The powerlaw exponent
values obtained are thus displayed in Tab. \ref{nh-gamma-tbl}.

If we pay attention to the real spectra when a disc is required, the
exponent values range from a weak dependency of 0.15 in GRO J1655-40
to a bigger one of 2.25 in SWIFT J1753.5-0127. The accuracy of these
exponent value is higher than 2.5$\sigma$ for the four sources
studied. If we now examine the values obtained with the \texttt{fff}
spectra, we note that the exponent is still positive, except for GRO
J1655-40, suggesting a possible degeneracy.  However, the $N_{\rm H}$
vs $\Gamma$ dependency is lower in the \texttt{fff} case, as shown in
in Tab. \ref{nh-gamma-tbl}.  The exponent values obtained in the real
case are always higher than those obtained in the \textit{fff} case by
a factor 2.5 to 4.7.  Examination of the same analysis performed on the whole
dataset (right hand side of Tab. \ref{nh-gamma-tbl}) gives similar
results.

\begin{table*}
\centering 
\caption{Result of the fit by a powerlaw plotted on
  Fig. \ref{gamma-nh}. Dashed line of Fig. \ref{gamma-nh}
  correspond to the ``only obs. with disc'' where dotted-dashed
  line refers to ``all obs.''. All uncertainties are $1\ \sigma$
  wide. These results are mainly consistent with a real $N_{\rm H}$
  versus $\Gamma$ correlation, however partly due to systematic
  deviation coming from spectral analysis. } 
\begin{tabular}{|c|c|c|c|c|}
\hline
&\multicolumn{4}{c|}{Model : $N_{\rm H}=\alpha\times \Gamma^\beta$}\\
\hline
\multirow{2}*{Source}&\multicolumn{2}{c|}{$\beta$ (only obs. with disc)} & \multicolumn{2}{c|}{$\beta$ (all obs.)}\\
&real & \textit{fff} & real &  \textit{fff} \\
GX 339-4&$0.47\pm0.05$ &$0.10\pm0.09$&$0.98\pm0.14$&$0.09\pm0.06$\\
XTE J1817-330&$0.85\pm0.32$&$0.35\pm0.14$&$1.11\pm0.39$&$0.14\pm0.10$\\
SWIFT J1753.5-0127&$2.25\pm0.38$&$0.86\pm0.35$&$1.16\pm0.40$&$0.15\pm0.09$ \\
GRO J1655-40&$0.15\pm0.06$&$-0.009\pm0.014$&$0.23\pm0.11$&$-0.01\pm0.014$\\
\hline
\label{nh-gamma-tbl}
\end{tabular}
\end{table*}

{\bf We then conclude that if a powerlaw is used as model to fit the
  high energy part of the spectrum, the value of $N_{\rm H}$ seems to
  vary during the evolution of the source, and \textit{Swift}/XRT is
  able to detect this evolution. As a result,  we suggest to keep
  $N_{\rm H}$ as a free parameter when fitting \textit{Swift}/XRT data
  with this model}\footnote{Of course, this will only be valid for
  high enough $N_{\rm H}$.}, and, we suggest to use the \texttt{fff}
method for further  $N_{\rm H}$ variation studies with other
observatories.

\subsection{High energy component fitted by comptonisation model.}

As the underlying processes resulting in the high energy tail are
still under debate, we also chose to examine if the previous variation
of $N_{\rm H}$ was still evident when using another model. We thus
chose {\sc xspec} \texttt{comptt} model, often used as a physically
motivated alternative compared to a powerlaw. But this model has the
characteristic to exhibit a low energy ``cut-off'' (Rayleigh-Jeans
part of the seed photons coming from the black-body emission), which
can greatly influence our previous analysis. If the ``true'' emission
process is actually not extended to low energy, and a powerlaw (which
by definition exhibits no low energy cutoff) is used, one expects the
absorption to play the role of an artificial cutoff. Hence the $N_{\rm
  H}$ value obtained when using a powerlaw will be higher than when
using \texttt{comptt}.

\begin{figure*}
\vbox to 220mm{\vfil 
\includegraphics[width=0.75\textwidth]{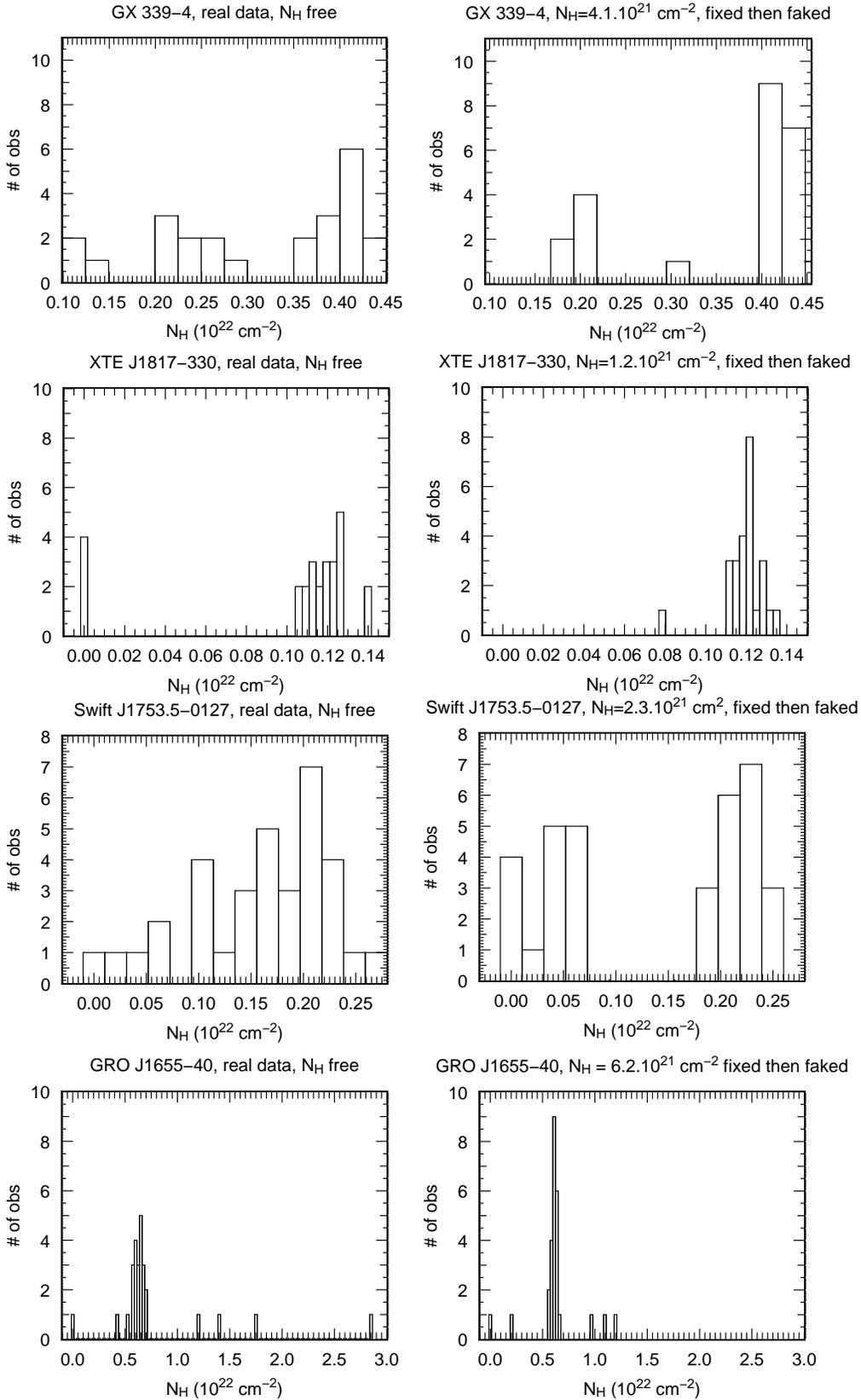}
 \caption{Distribution of $N_{\rm H}$ when high energy component in
   real and faked data are fitted with the \texttt{comptt} model.
} 
 \label{histo-nh-comptt}
\vfil
} 
\end{figure*}

\begin{figure*}
\vbox to 220mm{\vfil 
\includegraphics[width=0.75\textwidth]{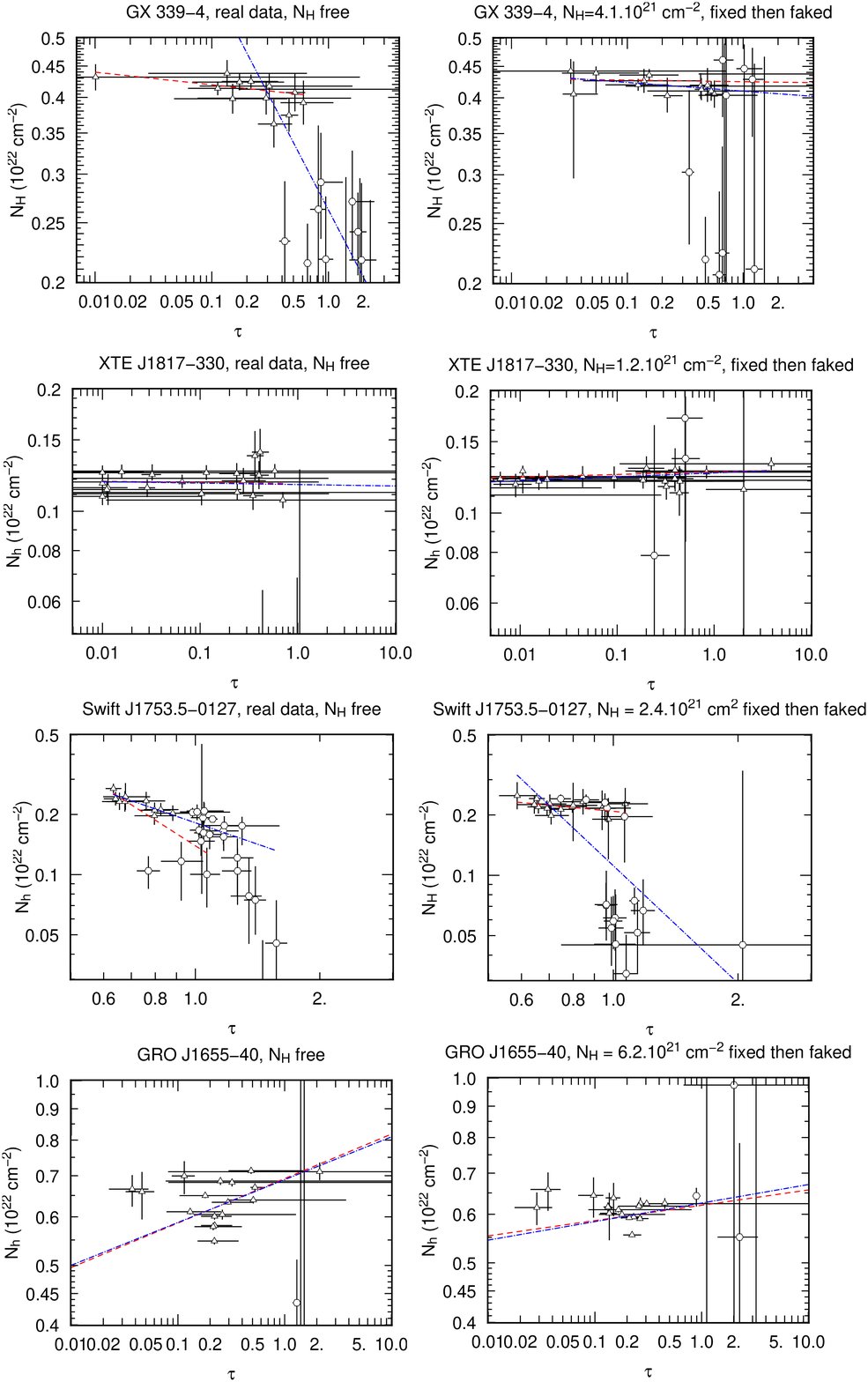}
 \caption{Column density variation during the transition in GX 339-4,
   XTE J1817-330, SWIFT J1753.5-0127 and GRO J1655-40 when high energy
   component in real and faked data are fitted with the
   \texttt{comptt} model. See Tab. \ref{nh-tau-tbl} for the fit
   results.  
} 
 \label{tau-nh-comptt}
\vfil
} 
\end{figure*}

Thus we reran the previous analysis with this model and the
results are displayed in Fig. \ref{histo-nh-comptt} and
\ref{tau-nh-comptt} and Tabs. \ref{nh-tau-tbl} and
\ref{nh-ks-test-tbl}.  Firstly, we can visually examine \textit{fff}
distributions obtained on right-hand side of
Fig. \ref{histo-nh-comptt}. We note that instead of being
normally distributed around the adopted weighted mean value of $N_{\rm
  H}$, the \textit{fff} spectra tend here to exhibit a second peak at 
low $N_{\rm H}$ for GX 339-4 (the six observations where $N_{\rm H}
\sim 2\times 10^{21}\ {\rm cm^{-2}}$) and SWIFT J1753.5-0127 (11
observations with $N_{\rm H} \sim 5\times 10^{20}\ {\rm
  cm^{-2}}$). This means that faking data with a constant $N_{\rm H}$
can, in certain cases when using the \texttt{comptt} model and if we
refit the same spectrum with a floating absorption, lead to quite
significant differences in the $N_{\rm H}$ fitted value. Moreover,
examining the KS test results (right hand side of
Tab. \ref{nh-ks-test-tbl}) gives a higher probability that real and
\textit{fff} distributions are the same when using \texttt{comptt}
compared to when using a powerlaw (except for GRO J1655-40). This effect is
particularly visible for SWIFT J1753.5-0127 where the KS probability
reaches 15 per cent.

\begin{table*}
\centering 
\caption{Results of the fit by a powerlaw plotted on Fig. \ref{tau-nh-comptt}.  Every terms have the same signification as Tab. \ref{nh-gamma-tbl}. Here, except with XTE J1753.5-0127, there seem to be less correlation between the value of the warm plasma optical depth  and the column density.}
\begin{tabular}{|c|c|c|c|c|}
\hline
&\multicolumn{4}{c|}{Model : $N_{\rm H}=\alpha\times \tau^\beta$}\\
\hline
\multirow{2}*{Source}&\multicolumn{2}{c|}{$\beta$ (only obs. with disc)} & \multicolumn{2}{c|}{$\beta$ (all obs.)}\\
&real & \textit{fff} & real &  \textit{fff} \\
GX 339-4&$-0.02\pm0.01$ &$-0.025\pm0.008$&$-0.37\pm0.30$&$-0.009\pm0.014$\\
XTE J1817-330&$-0.003\pm0.009$&$0.008\pm0.003$&$-0.004\pm0.006$&$0.010\pm0.003$\\
SWIFT J1753.5-0.127&$-0.51\pm0.16$&$-0.19\pm0.14$&$-0.71\pm0.14$&$-1.9\pm0.3$ \\
GRO J1655-40&$0.072\pm0.031$ &$0.025\pm0.039$ &$0.070\pm0.030$ & $0.030\pm0.032$\\
\hline
\label{nh-tau-tbl}
\end{tabular}
\end{table*}

In order to probe if any correlation between
the state and the value of $N_{\rm H}$ remains, we used the plasma optical
depth value, $\tau$, as an indicator equivalent to the photon index
(since $\Gamma$ depends on the compton parameter $y$, where
$y=kT_e/(m_ec^2)\times max(\tau,\tau^2)$ and $T_e$ is fixed at 50
keV). Fig. \ref{tau-nh-comptt} illustrates the possible $N_{\rm H}$
vs $\tau$ dependency, and fits by powerlaws have been performed as in
the previous section. Values of the exponent obtained for observations
where a disc is required are consistent with no variation of $N_{\rm
  H}$ vs $\tau$ in GX 339-4, XTE J1817-330 and GRO J1655-40. It is
either due to the fact that the value of the exponent is similar
between the real and the \textit{fff} spectra (GX 339-4 case), or
because its $1\sigma$ error for real spectra is as large as the value
obtained (XTE J1817-330 case). GRO J1655-40 seems even to show inverse
behaviour - the value of the hydrogen column density seems indeed to
correlate with the value of the plasma optical depth. On the contrary,
SWIFT J1753.5-0127 still seems to exhibit significant decrease of
$N_{\rm H}$ when $\tau$ increase.

Considering the complete dataset (dot-dashed curve in
Fig. \ref{tau-nh-comptt} and right hand side of Tab. \ref{nh-tau-tbl})
returns similar results, except for SWIFT J1753.5-0127, where the
bimodality of \textit{fff} distribution is quite noticeable, forming
two separated cluster of points (lower right graph of Fig. \ref{nh-tau-tbl}).

\begin{table}
\centering 
\caption{Results given by a KS test probing the probability that the
  underlying distributions of $N_{\rm H}$ obtained in
  Fig. \ref{histo-nh-po} are one and the same for the real and the
  faked spectra.
}
\begin{tabular}{|c|c|c|c|c}
\hline
&\multicolumn{2}{c|}{\texttt{powerlaw}} &\multicolumn{2}{c|}{\texttt{comptt}} \\
\hline
KS test: &value&prob.&value&prob.\\
\hline
GX 339-4&0.34&0.048&0.375&0.051\\
XTE J1817-330&0.41&0.016&0.33&0.09\\
SWIFT J1753.5-0.127&0.73&$7.1\times 10^{-9}$&0.26&0.15\\
GRO J1655-40&0.28&0.24&0.41&0.017\\
\hline
\label{nh-ks-test-tbl}
\end{tabular}
\end{table}

Thus, when using \texttt{comptt} to fit the high energy
tail of the spectra in \textit{Swift}/XRT data, there is less evidence
that the absorption column density is varying.  As a result one should view the
fitted value of $N_{\rm H}$ with caution. {\bf However, as the
  statistical errors obtained on $\tau$ in this study are quite large,
  it is difficult 
  to draw definite conclusions about the $N_{\rm H}$ vs $\tau$
  dependency.} The simultaneous use of an instrument which gives more accurate
spectra at higher energy together with our prescribed \texttt{fff}
method could help to probe any  $N_{\rm H}$ vs $\tau$ variation. {\bf
  As a general conclusion, we suggest to leave the absorption free to
  vary when fitting \textit{Swift}/XRT spectra}.

 \section{A recessing disc when $L_{\rm bol}<10^{-2}\ L_{\rm Edd}$ in
   XTE J1817-330?}\label{subsub-nh-disk}

Although we have just investigated in which cases one has to let the
value of $N_{\rm H}$ float during the fitting process, we will examine
in this section what is its impact on the disc properties is,
especially its geometry. Thus, we will still consider on the one hand,
the case when $N_{\rm H}$ is free to vary, and on the other hand, when
it is frozen to the value obtained at high luminosity (see
Sect. \ref{subsec-influ-nh} for those values). This work focuses on
the \textit{Swift}-XRT observations of XTE J1817-330 which appear to
show significant changes in the fitted disc properties during its
decline.

Fixing the value of $N_{\rm H}$, or leaving it free during the fitting
process has a major impact on the possible detection of a disc
component. However, by comparing the effect of adding a \texttt{diskbb} model,
it seems that in both cases the fit is improved at high luminosity. In
the case of an $N_{\rm H}$ value fixed to the one obtained at high
luminosities (as opposed to the case where $N_{\rm H}$ is free to
vary), taking the example of observation number 13, the $\chi^2$ drops
from $13720$ to $386.2$  (respectively $2281.9$ to $380.9$) when
adding the thermal component.

We now focus particularly on the last four observations of XTE
J1817-330 and especially observation number 20. The goodness of fit
summary is shown in Tab. \ref{chi2-ryk}. Taking into account the
degrees of freedom, the situation is less clear, it can show
that a simple absorbed powerlaw already gives a relative good fit when
the $N_{\rm H}$ is free (reduced $\chi^2=1.08$) and adding a
\texttt{diskbb} in this case suggests an overfitting of the data (the
F-test probability is higher than 3 per cent). On the contrary, when the
$N_{\rm H}$ is fixed to the value obtained at high luminosity, as the
absorption in this latter case is higher, adding an extra component at
low energies, like a \texttt{diskbb}, will be required (see
Tab. \ref{chi2-ryk}). This effect is due to the fact that, as we
demonstrated in previous sections, when using a powerlaw for the high
energy component, the absorption seems to slightly decrease when
entering the harder states. Thus an extra component at low energy such
as a disc would not be required anymore. On the contrary, if we fix
the $N_{\rm H}$ to a high value, then a disc component would be
necessary to compensate the flux lost by the absorption.

In the next sections, considering that this effect could affect the
possible detection of a disc component, we adopted the following
strategy when analysing \textit{Swift} data: 

\begin{itemize}
\item For easier comparisons and consistency with earlier results
  given in the bibliography, we fitted the high energy part of the
  spectra with a powerlaw.

\item If ever the spectrum is well fitted by a simple absorbed
  powerlaw (threshold chosen: $\chi^2/\nu<1.2$), or if adding a disc
  component to the latter model gives a high  F-test probability
  (hereafter, $>10^{-2}$), then no disc component is added.

\item On the contrary, when an extra component is necessary, we study
  the evolution of the parameters and we allowed the absorption to
  vary (unless it is explicitly mentioned as for XTE J1817-330 case
  where both varying and fixed $N_{\rm H}$ case were studied).

\end{itemize}

We can also compare our results with those obtained by R07. For
example, in observation number 19 we obtain $\chi^2/\nu=18.33/23$ with
a simple absorbed powerlaw only. The absorption value obtained in this
case is quite low ($N_{\rm H}=7.2\pm 5\,\times 10^{20}\ {\rm
  cm^{-2}}$), however consistent with earlier Chandra observation
\citep{mi06c}. Even if we set the $N_{\rm H}$ value to $1.2\ \times
10^{21}\ {\rm cm^{-2}}$, we obtain $\chi^2/\nu=20.02/24$, which is
already sufficient to model the spectrum properly (the addition of a
\texttt{diskbb} does not improve the fit). It is not clear why R07
added an extra component in that observation. We have similar concerns
for obs. 20 and 22. Examining the $\chi^2$ values obtained (see
Tab. \ref{chi2-ryk}), one can see that adding a disc component is not
necessary for the four last observations except, as noted above, in
observation number 20. Note also that in obs. 20 the spectrum is only
made of 27 channel bins. Performing a F-test between both models (with
or without disc) gives a probability of $2\times 10^{-3}$, which is
still quite high and still cast doubt on the reliability of adding
such thermal component. This disc measurement was however kept for
this section's study.

The variation of the hydrogen column density value has also been
reported in previous papers involving observations with \textit{Swift}
(see e.g. \citealt{bro06} for the GRO J1655-40 outburst\footnote{Based
  on the same data set used in this paper.}) for objects in their soft
state but also in their hard states (\citealt{ost96} in V404-Cyg). We
also saw in Sect. \ref{sec-data-red} that ASCA/GIS observations of GX
339-4 can be interpreted in the same framework and lead to different
conclusions concerning the absence or the presence of a thermal
component whether the value of $N_{\rm H}$ is fixed or not.

We now focus on the remaining observations of XTE J1817-330 at higher
luminosity, (observations number two to 17 and even 20 if $N_{\rm H}$
is fixed). When a disc is required, one can track the evolution of the
source as its flux is monotonically declining from the first to the
last observation (see also \citealp{gd08}). As noticed in \cite{gd08}
by contemporaneous observations with RXTE, in the hardness-intensity
diagram the source is fading from the high soft state (HS) to the low
hard state (LH) by transiting through the lower intermediate state
branch (IM) \citep{homan05}. Note that observations number 15, 16 and
17, of particular interest as we will see, are located in this latter
branch.

\begin{figure*}
\vbox to 220mm{\vfil 
\includegraphics[width=0.75\textwidth]{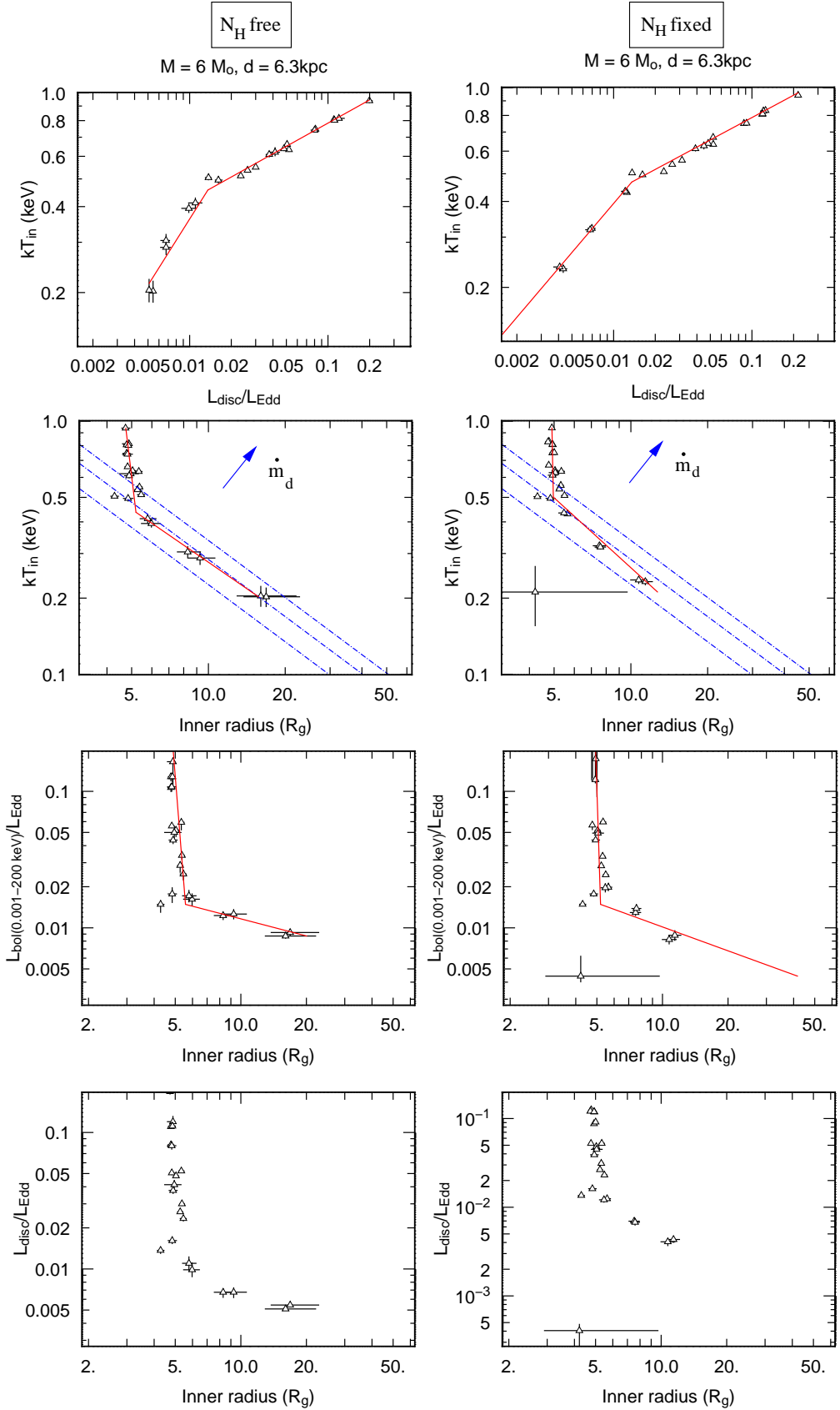}
\caption{Variation of the disc properties during the decline of XTE
  J1817-330 when fitting the hard component with a powerlaw. The
  $kT_{\rm in}$ vs $L_{\rm disc}/L_{\rm Edd}$, $kT_{\rm in}$ vs
  $R_{\rm in}/\rm R_{\rm g}$ and $L_{\rm bol}/L_{\rm Edd}$ vs $R_{\rm
    in}/\rm R_{\rm g}$ data sets have been fitted by a broken powerlaw
  (solid lines, see text and Tab. \ref{var-par-1817-tbl} for the
  parameters of the fit). The dotted dashed lines plotted in the
  second row of graphs are showing constant $\dot{m}_{\rm d}$
  profile. From upper right to lower left: $\dot{m}_{\rm
    d}=0.002,\ 0.005,\ 0.01$ with $\eta_{\rm Edd}=0.1$.} \label{1817-var}
\vfil
} 
\end{figure*}

\begin{figure*}
\vbox to 220mm{\vfil 
\includegraphics[width=0.75\textwidth]{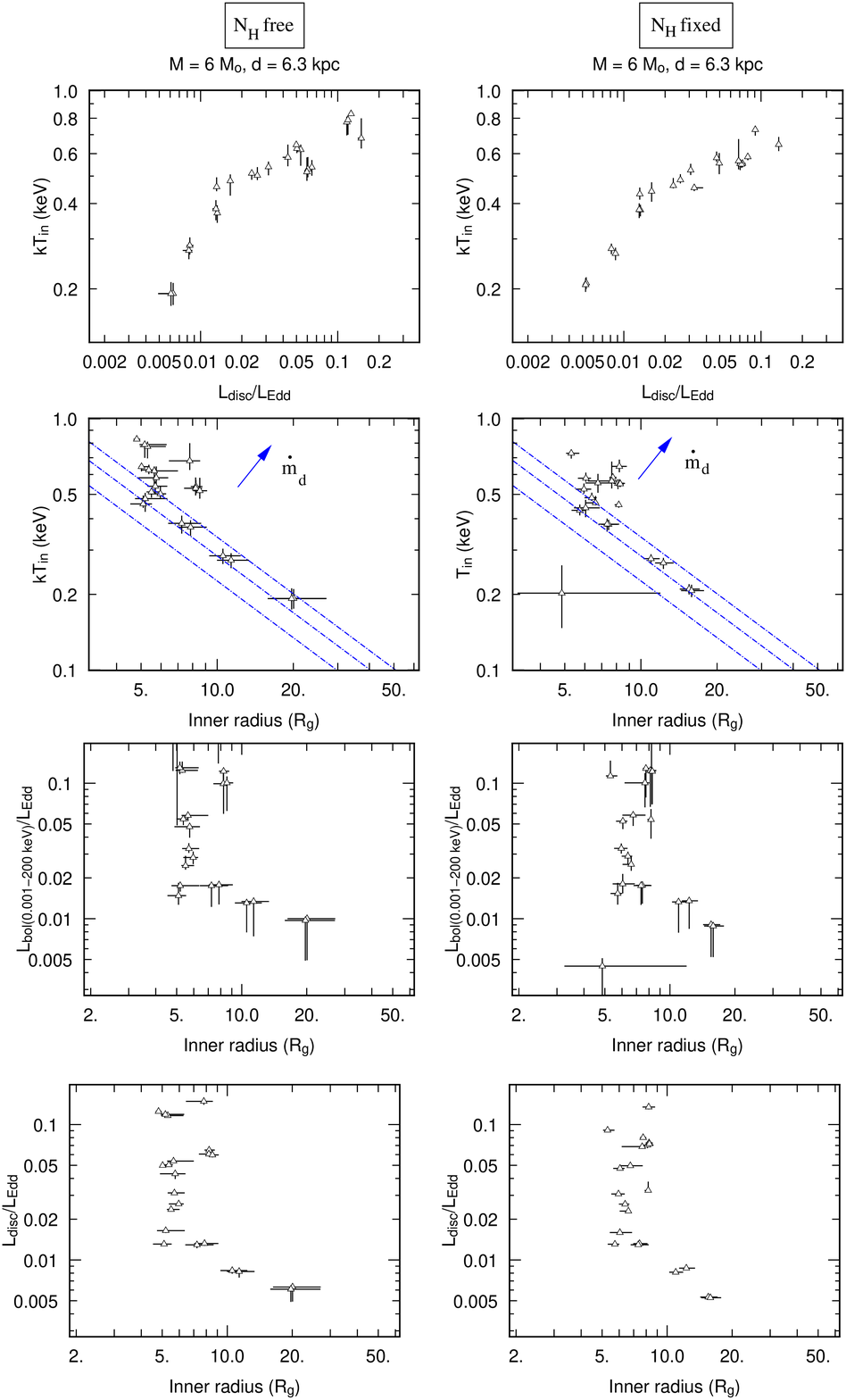}
\caption{Variation of the disc properties during the decline of XTE J1817-330 when fitting the hard component with the \texttt{comptt} model. Here again the beginning of a disc recession seems to be needed. Even if the behaviour at high luminosity is less constant. Same values as in Fig. \ref{1817-var} for $\dot{m}_{\rm d}$ in the second graph row.} \label{1817-var-comptt}
\vfil
} 
\end{figure*}

The whole detailed evolution of the disc properties is displayed in
Fig. \ref{1817-var} when using a simple powerlaw as a model to fit the
hard component of the spectra, and in Fig. \ref{1817-var-comptt} when
using \texttt{comptt} for the same purpose. As mentioned before, we
also examined both the cases where $N_{\rm H}$ is fixed or not: as
there are 4 observables ($L_{\rm disc}$, $L_{\rm bol}$, $R_{\rm in}$,
$kT_{\rm in}$), there are in theory six 2-dimensional plot
combinations. However, for simplicity, we restricted this number to
the 4 most important for probing the source evolution and we discarded
the $L_{\rm bol}$ vs $L_{\rm \rm disc}$ and the $kT_{\rm in}$ vs
$L_{\rm bol}$ plots.

Examination of the first row of graphs in Fig. \ref{1817-var}
($kT_{\rm in}$ vs $L_{\rm disc}$) seems to clearly show an obvious trend:
where at high luminosity the temperature seems to decrease
monotonically as a unique powerlaw, the last three ``double points''
(there was two GTIs for each of those three observations) shows a
clear drop in temperature. We then tried to fit this $kT_{\rm in}$ vs
$L_{\rm disc}$ relationship with two models: first using a simple
powerlaw, and second with a broken powerlaw. The fit
algorithm uses the standard merit function described in \cite{press92}, \textsection\ 6.7. It does take into account both the errors on the Y (inner
temperature) and X (luminosity) axis: the
  principle of this fitting proceedure is described in details in \cite{press92},  \textsection\  15.3, ``Straight-Line Data with Errors in Both Coordinates''. This algorithm was also used when fitting the $kT_{\rm in}$ vs $R_{\rm in}$ and $L_{\rm bol}$ vs $R_{\rm in}$ relationships.

 The results are given in the first set of rows of
 Tab. \ref{var-par-1817-tbl}: first, a simple powerlaw to model those
 data give a very bad fit, whenever the value of $N_{\rm H}$ is fixed
 or not ($\chi^2_{\rm spl}/\nu \sim 5.0$ or $\sim 10.$) . Using a
 broken powerlaw significantly improves the fit as it drops to a value
 of $1.15$ ($N_{\rm H}$ free) or $2.3$ ($N_{\rm H}$ fixed). Second,
 the upper flux powerlaw index (0.27 or 0.26$\pm0.02$) is very close
 to the value expected for a disc emitting with a constant inner
 radius ($kT_{\rm in}\propto L_{\rm disc}^{0.25}$). By comparison, R07
 obtained a powerlaw index equal to $0.233\pm0.006$. It is interesting
 to note that the second powerlaw index found is quite steep and close
 to $0.77$ ($N_{\rm H}$ free) or 0.6 ($N_{\rm H}$ fixed) at lower
 flux. The errors on those values, although bigger, are still consistent
 with a break which would occur at around $0.014\ L_{\rm Edd}$ (with
 orbital parameters and distance given in
 Tab. \ref{tab-mass-dist-angle}). As the temperature seems to suddenly
 decrease for a rather constant disc luminosity, one expects its
 normalisation and hence the internal radius to increase in order to
 compensate for this fall. This is what is examined in the six lower
 panels of Fig. \ref{1817-var}. In observations 16 and 17, the
 internal radius values double and triple from those measured at
 higher luminosities. 

\begin{table*}
\centering 
\caption{$\chi^2$ values when fitting the 4 last \textit{Swift}
  observations of XTE J1817-330 with or without a disc component and
  using the latest response files. Even when $N_{\rm H}$ is fixed to
  $0.12\times 10^{22}\ {\rm cm^{-2}}$, the reduced $\chi^2$ is quite
  good enough (i.e. $<1.15$) to model the data with a simple powerlaw
  in observations 19, 21 and 22. Adding a disc component would overfit
  the data in this case. Note that the values of the f-test
  probability are just informative as its use is controversial for
  such small data sets (\citealp{protassov02}).}

\begin{tabular}{|c|c|c|c|c|}
\hline
 & \multicolumn{4}{c|}{$\chi^2/\nu$}\\
\hline
Model& Obs. 19& Obs. 20& Obs. 21& Obs. 22\\
\hline
wabs$\times$po&$\frac{18.33}{23}=0.80$&$\frac{25.89}{24}=1.08$&$\frac{27.84}{31}=0.90$&$\frac{35.40}{38}=0.93$\\
\\
wabs$\times$(diskbb+po)&$\frac{17.53}{21}=0.83$&$\frac{19.16}{22}=0.87$&$\frac{26.10}{29}=0.90$&$\frac{28.63}{36}=0.80$\\
\hline
ftest probability&0.63&0.036&0.39&0.022\\
\hline
\hline
wabs$\times$po ($N_{\rm H}$ fixed)&$\frac{20.02}{24}=0.83$&$\frac{38.41}{25}=1.54$&$\frac{36.82}{32}=1.15$&$\frac{39.81}{39}=1.02$\\
\\
wabs$\times$(diskbb+po) ($N_{\rm H}$ fixed)&$\frac{18.28}{22}=0.83$&$\frac{22.27}{23}=0.97$&$\frac{27.88}{30}=0.93$&$\frac{32.85}{37}=0.89$\\
\hline
ftest probability&0.37&0.002&0.015&0.038\\
\hline
\label{chi2-ryk}
\end{tabular}
\end{table*}

\begin{table*}
\centering 
\caption{Fit results of Fig. \ref{1817-var}.}

\begin{tabular}{|c|c|c|c|c|c|c|}
\hline
 & \multicolumn{5}{c|}{$kT_{\rm in}=N\times (L_{\rm disc}/(L_{\rm Edd}\times b))^{\alpha_1}$ when $L_{\rm disc}/L_{\rm Edd}> b$}&\multirow{2}*{$kT_{\rm in}=N \times (L_{\rm disc}/L_{\rm Edd})^{\alpha_{\rm spl}}$}\\
&\multicolumn{5}{c|}{$kT_{\rm in}=N\times (L_{\rm disc}/(L_{\rm Edd}\times b))^{\alpha_2}$ when $L_{\rm disc}/L_{\rm Edd}< b$}&\\
\hline
&$b$&$N$&$\alpha_1$&$\alpha_2$&$\chi_{\rm bknpl}^2/\nu$&$\chi_{\rm spl}^2/\nu$\\
\hline
$N_{\rm H}$ free&$0.014\pm0.001$&$0.46\pm0.021$&$0.27\pm0.018$&$0.77\pm0.15$&$22.9/20$&$109.2/22$\\
$N_{\rm H}$ fixed&$0.014\pm0.001$&$0.47\pm0.016$&$0.26\pm0.018$&$0.57\pm0.1$&$46.04/20$&$214/22$\\
\hline
\hline
 & \multicolumn{5}{c|}{$R_{\rm in}/\rm R_{\rm g}=N\times (L_{\rm bol}/(L_{\rm Edd}\times b))^{\alpha_1}$ when $L_{\rm bol}/L_{\rm Edd}> b$}&\multirow{2}*{$R_{\rm in}/\rm R_{\rm g}=N \times (L_{\rm bol}/L_{\rm Edd})^{\alpha_{\rm spl}}$}\\
&\multicolumn{5}{c|}{$R_{\rm in}/\rm R_{\rm g}=N\times (L_{\rm bol}/(L_{\rm Edd}\times b))^{\alpha_2}$ when $L_{\rm bol}/L_{\rm Edd}< b$}&\\
\hline
&$b$&$N$&$\alpha_1$&$\alpha_2$&$\chi_{\rm bknpl}^2/\nu$&$\chi_{\rm spl}^2/\nu$\\
\hline
$N_{\rm H}$ free&$0.015\pm0.003$&$5.56\pm0.32$&$-0.05\pm0.03$&$-2.4\pm1.4$&$55.8/20$&$152/22$\\
$N_{\rm H}$ fixed&$0.015\pm0.002$&$5.20\pm0.30$&$-0.02\pm0.02$&$-1.72\pm0.6$&$37.09/20$&$299/22$\\
\hline
\hline
 & \multicolumn{5}{c|}{$R_{\rm in}/\rm R_{\rm g}=N\times (kT_{\rm in}/b)^{\alpha_1}$ when $kT_{\rm in}> b$}&\multirow{2}*{$R_{\rm in}/\rm R_{\rm g}=N \times (kT_{\rm in})^{\alpha_{\rm spl}}$}\\
&\multicolumn{5}{c|}{$R_{\rm in}/\rm R_{\rm g}=N\times (kT_{\rm in}/b)^{\alpha_2}$ when $kT_{\rm in}< b$}&\\
\hline
&$b$&$N$&$\alpha_1$&$\alpha_2$&$\chi_{\rm bknpl}^2/\nu$&$\chi_{\rm spl}^2/\nu$\\
\hline
$N_{\rm H}$ free&$0.44\pm0.10$&$5.20\pm0.60$&$-0.12\pm0.15$&$-1.44\pm0.7$&$82.7/20$&$107.34/22$\\
$N_{\rm H}$ fixed&$0.50\pm0.03$&$4.96\pm0.57$&$-0.02\pm0.09$&$-0.99\pm0.2$&$47.7/20$&$175.6/22$\\
\hline
\label{var-par-1817-tbl}
\end{tabular}
\end{table*}

The second row of Fig. \ref{1817-var} shows the variations of $kT_{\rm
  in}$ as a function of $R_{\rm in}$. A broken powerlaw has been used
again to fit the data. Since for a pure multicolour disc each ring is dissipating the
gravitational energy available in radiation as it is optically thick,
it can be used as a direct probe of the accretion rate.  In that case
we have:

\begin{equation}
T_{\rm in}=\left(\frac{3GM\dot{M}_{\rm d}}{8\pi R_{\rm in}^3\sigma_{\rm B}}\right)^{1/4},
\end{equation}
(see e.g \citealp{belloni97}), where $\dot{M}_{\rm d}$ is the mass
accretion rate flowing through the disc only. This supposes that 100
per cent
of the mass coming from the donor star is accreted entirely through
the disc until $R_{\rm in}$.  Therefore, renormalising the same
formula to Eddington accretion rate and 
using $L_{\rm Edd}=\eta_{\rm Edd}\dot{M}_{\rm Edd}c^2$, we obtain:

\begin{equation}
\frac{kT_{\rm in}}{1\ \mathrm{keV}}=5.31\left(\frac{M}{\mathrm{M_{\odot}}}\right)^{-1/4}\left(\frac{\dot{m}_{\rm d}}{\eta_{\rm Edd}}\right)^{1/4}\left(\frac{R_{\rm in}}{\rm R_{\rm g}}\right)^{-3/4},\label{eq-kt-mdot-rin} 
\end{equation}
where $\dot{m}_{\rm d}=\dot{M}_{\rm d}/\dot{M}_{\rm Edd}$. { \bf Note
  that  $\bf \dot{m}_{\bf d}\neq \dot{m}_{\bf total}$ in the general
  case (see for example \citealp{vklis01} for evidences of this
  discrepancy in LMXB)}.

Hence, parallel lines with a slope of (-3/4) can be drawn in a log-log
temperature versus radius diagram and the normalisation value only
depends on the values of the mass, the radiative efficiency at
Eddington luminosity and the accretion rate. This has been done in the
second rows of Fig. \ref{1817-var} and \ref{1817-var-comptt}. Below
the apparent break in the relationship, we observe that the disc
seems to decrease its accretion efficiency by increasing its radius,
keeping its accretion rate constant, especially when $N_{\rm H}$ 
is kept free. This is further supported by the value of the slope
given by the fit even if the errors on it are quite high
($-0.70\pm0.33$ consistent with -3/4). By taking $\eta_{\rm Edd}=0.1$,
we conclude then that the accretion rate is close to 0.005 Eddington
($\pm0.0025$ with the large uncertainties on the BH mass). This effect
also seems to hold when using a \texttt{comptt} (see second row of
Fig. \ref{1817-var-comptt}). Using $L_{disc}=2\pi R_{\rm
  in}^2\sigma_{\rm B}T_{\rm in}^4$, we can easily calculate the
value of the accretion rate through the disc\footnote{Note that the
  equation number \ref{eq-mdot-ldisc} is obtained in the framework of
  the multicolour disc model. One known limitation is that it gives a
  quite high accretion efficiency when the radius approach the ISCO,
  as in that case $\eta(R_{\rm in})_{\texttt{diskbb}}=3/2({\rm R_{\rm
      g}}/R_{\rm in})=0.25\neq 0.06$ when $R_{\rm in}= 6\ \rm R_{\rm
    g}$ for example.}, i.e.:

\begin{equation}
\dot{m}_d=\frac{2}{3}\eta_{\rm Edd}\left(\frac{L_{\rm disc}}{L_{\rm
    Edd}}\right)\left(\frac{R_{\rm in}}{\rm R_{\rm
    g}}\right).\label{eq-mdot-ldisc}
\end{equation}

Moreover, as variations in the mass ratios of BHBs are far less than
those on accretion rates ($10^8$), one can set a typical mass value
(e.g $8\mathrm{M_{\odot}}$) and Eddington accretion efficiency (e.g
$\eta_{\rm Edd}=0.1$) in order to estimate the evolution of the
accretion rate among different sources. This will be the subject of
Section \ref{sec-multi}.

The beginning of the disc recession can also be noticed in the two
last rows of graphs of Fig. \ref{1817-var} and \ref{1817-var-comptt}
and in Tab. \ref{var-par-1817-tbl} for the results of the fit by a
broken powerlaw: the crucial transition observed in the geometry for
this source seems then to occur at the same bolometric luminosity
level ($\sim 0.015 L_{\rm Edd}$). This is due to the fact that until
observation number 15, the disc contributes to more than 90 per cent of the
overall luminosity (it drops to 58 per cent in observation 17).

We also compared and included with our results the XMM observation of
the source analysed by \cite[S07]{sala07}. For a better comparison
with our data processing manner, we used the fit results that S07
obtained with an absorbed \texttt{powerlaw+diskbb}.  It appears that
the disc normalisation value is still in the range we obtain with
\textit{Swift} data in the highest fluxes state (as $K_{\rm
  disc}\simeq 2000$) although the value of the photon index differs
quite a lot when the value of $N_{\rm H}$ is tied to a value of
$1.5\times 10^{21} {\rm cm^{-2}}$. With $\Gamma\simeq 1.6$, the XMM-Newton
observation seems to exhibit a non-recessed disc in a low-hard
state, similarly to the observations analysed by M06 on GX
339-4. However, if we focus on the value of the fluxes obtained using
the fit parameters values given in S07 with $N_{\rm H}$ frozen, it
results in an unabsorbed disc flux of $F_{\rm disc}\simeq 1.1\ \times
10^{-8}\ \mathrm{erg\ cm^{-2}\ s^{-1}}$ and a high absorbed
powerlaw flux of $F_{\rm pl}=2.4\ \times
10^{-7}\ \mathrm{erg\ cm^{-2}\ s^{-1}}$ in the 0.02-200 keV
range. Thus in terms of Eddington luminosity it gives $L_{\rm
  disc}\sim 0.060\ L_{\rm Edd}$ but $L_{\rm bol}\sim 1.3\ L_{\rm Edd}$
($M=6\mathrm{M_{\odot}}$, $d=6.3\rm\ kpc$, see
Tab. \ref{tab-mass-dist-angle}). Unless one cannot rule out the
possibility that the source lies in our vicinity (i.e $d\sim
1\rm\ kpc$, in that case $L_{\rm hard}\sim 0.012\ L_{\rm Edd}$), the
overall luminosity is quite high. 

We can also note that S07 obtained a much higher photon index
($\Gamma\simeq 2.3$) when letting the value of $N_{\rm H}$ free to
vary, but the high energy flux is still completely dominating the
spectrum. Such a high bolometric flux is quite puzzling as the closest
\textit{Swift}/XRT observations of the source taken 6 days before and
two days after (7th and 15th March 2006, corresponding to ObsIDs
number 03 and 04) are one order of magnitude less luminous ($L_{\rm
  bol}\sim 0.13$ and $0.11 L_{\rm Edd}$ respectively).  We note
however that using \texttt{comptt} to model the high energy part of
the spectrum gives a more consistent bolometric luminosity value, as
in that case $L_{\rm bol}\sim 0.06 L_{\rm Edd}$. The discrepancy could
come from the fact that the $\chi^2$ value is far better with this
latter model rather than when using a simple powerlaw
($\chi^2/\nu=1.19$ instead of 1.45). However, in spite of those
drawbacks, for the consistency and comparisons among the different
sources and observation, we kept these results given by the fit with a
powerlaw (with $N_{\rm H}$ free).

 In the case that XTE J1817-330 lies at $1\rm kpc$, we would also
 obtain $R_{\rm in}\sim 0.77\pm 0.05 \rm\ R_{\rm g}<R_{\rm S}$ in the
 highest state of SWIFT data (as estimated from the first five
 observations).  This casts doubt on such a low value for the distance. In order
 to get at least $R_{\rm in}\sim R_{\rm S}$, one has to set a distance
 of $2.6\ \rm kpc$ and we thus took this model dependant value as
 minimal distance to XTE J1817-330 for the following multi-source
 study. Moreover, if we choose $d=1kpc$, the beginning of the disc recession
observed in SWIFT data would then occur at about $4\times
10^{-4}\ L_{\rm Edd}$.

\section{Multi-sources comparisons}
\label{sec-multi}

\begin{figure*}
\includegraphics[width=1.0\textwidth]{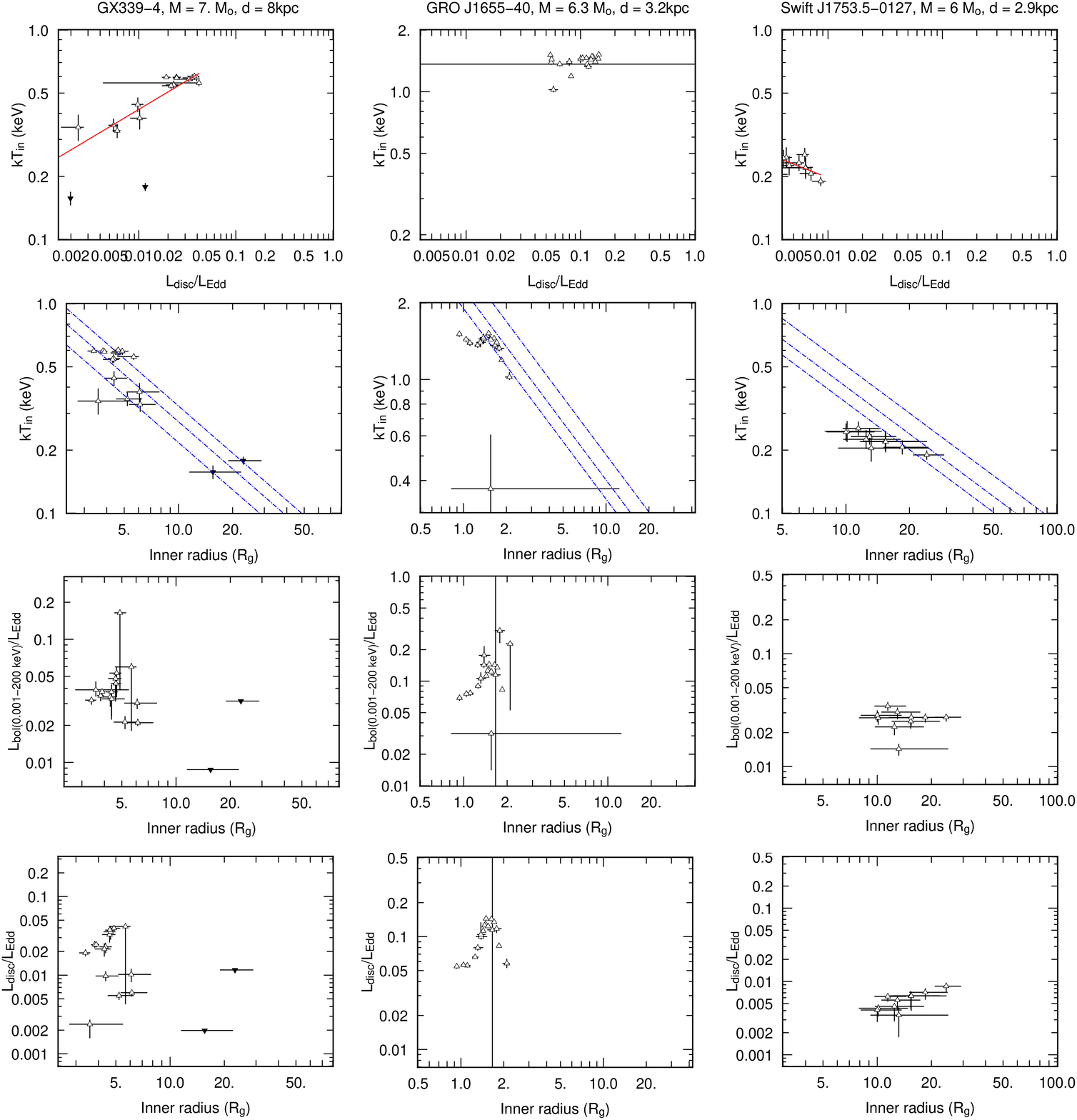}
\caption{ Variation of the disc properties of GX 339-4 (first column),
  GRO J1655-40 (second column) and SWIFT J1753.5-0127 (third column) as
  recorded by SWIFT. Note that in the case of GRO J1655-40, the source
  is in outburst whereas it is mainly in the decline for both
  others. On top left panel, fit of the cloud of point by a powerlaw
  gives an index of $\sim 0.10$. For SWIFT J1753.5-0127, this index is
  found to be negative ($\sim -0.21$): the disc luminosity decrease
  when the temperature increase. Concerning the $kT_{\rm in}$ vs
  $R_{\rm in}$ plots, same coding in $\dot{m}_{\rm d}$ as in
  Fig. \ref{1817-var} has been used. For GX 339-4, both points
  corresponding to the observations analysed by T08
  (\textcolor{black}{$\blacktriangledown$} symbols)
  with a simple absorbed \texttt{diskbb+powerlaw} have been
  added.}\label{all-sources-var-swift} 
\end{figure*}

While XTE J1817-330 is one of the best-studied examples, we need to
see how typical its behaviour is for BHBs. We first extended the
previous study to the other available SWIFT data of candidate
black holes, i.e GX 339-4, SWIFT J1753.5-0127 and GRO
J1655-40. According to the results obtained in
Sect. \ref{subsec-influ-nh}, we let the value of $N_{\rm H}$ free to
vary and used an absorbed \texttt{powerlaw} (and a \texttt{diskbb}
when needed) to model the spectra. Observations where a thermal
component is required are plotted in
Fig. \ref{all-sources-var-swift}. 

Contrary to the case of XTE J1817-330, there is less systematic
evolution of the disk radii. Indeed, if a faint increasing of its
value seems to occur at low luminosity in GX 339-4, the statistics are
not sufficient to draw any conclusions. The addition of the broadband
analysis of lower luminosity states led by \cite{tomsick08} (T08)
changes this a bit.  Indeed, when using reflection models, T08 seem to
obtain a geometry that does not evolve through the hard state (the
solid angle value remains constant and high enough). On the contrary,
the normalisation of the multicolour disc tends to exhibit quite high
value, whatever the high energy model chosen. For purpose of simpler
comparison between sources, we included the results coming from their
modelling by a simple absorbed \texttt{diskbb+powerlaw}.  
Moreover, the evolution of the bolometric luminosity is small (changes
by a factor of $\sim 5$ only) and thus further observations of the
decline would be necessary.  GRO J1655-40 exhibits unusual behaviour;
although the fitted radii change by a factor of two, the relationship
between the temperature and the disc luminosity is not monotonic but
rather quite complex. SWIFT J1753.5-0127 even shows a disc that tends
to get hotter when its luminosity decreases, with a higher internal
radius in more luminous state. However, we were limited by the
statistics and narrow range in luminosity sampled.

We then compared the whole set of sources in a single study. This has
been done by overplotting on the same graph the previous results, and
adding the few points obtained at lower luminosities. These latter
points mainly come from the low efficient BHB XTE J1118+480. However,
other recent or older observations of the previous sources (T08 for GX
339-4 and S07 for XTE J1817-330 as mentioned before) have been
added. The overall results are displayed in
Figs. \ref{fig-kt-ldisc-tout},  \ref{fig-rayon-lbol-tout},
\ref{fig-rayon-ldisc-tout}, \ref{fig-rayon-kt-tout},
\ref{fig-rayon-gamma-tout}, \ref{fig-lbol-mdot} and \ref{fig-dfld}.

\subsubsection*{Mass, distance and inclination angle uncertainties}

The uncertainties plotted directly on the graphs are the statistical
errors arising from the fits. We also attempted to estimate the
systematic uncertainties due to the the orbital parameters (see
Tab. \ref{tab-mass-dist-angle}). For simplification, we computed the
lower and upper boundaries by taking the limit values in masses,
distances and angles. For example, as the disc luminosity goes with
$L_{\rm disc}/L_{\rm Edd}\propto F_{\rm disc}d^2/(M\times \cos(i))$,
all the disc luminosities will be shifted by a factor $d_{\rm
  max}^2/(M_{\rm min}\times\cos(i_{\rm max}))$ if one wants to obtain the
highest value of $L_{\rm disc}/L_{\rm Edd}$.

 This corresponds to a vector in Fig. \ref{fig-kt-ldisc-tout}. We note, however,
 that this method is somewhat overestimating the shifts in the
 corresponding diagrams as there may be strong correlation among
 those parameters. As the mass is often determined via the mass
 function, in that case $M\times \sin^3(i)$ must remain constant when
 evaluating the uncertainties.  In addition, since for the recently
 discovered black-hole candidates 
(J1817-330 and J1753.5-0127) there is no proper orbital measures yet,
we have set quite large uncertainties.

\subsection{\rm Disc geometry evolution}

\begin{figure}
\includegraphics[width=0.5\textwidth]{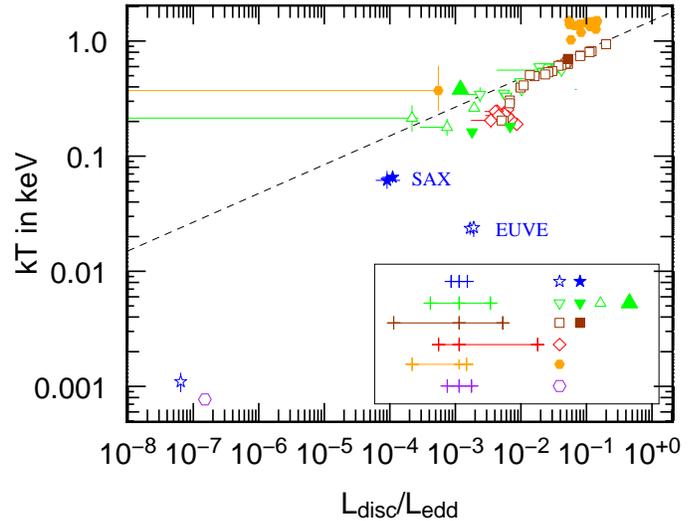}
  \caption{$kT_{\rm in}$ versus $L_{\rm disc}/L_{\rm Edd}$. The black
    dashed correspond to $T_{\rm in}\propto (L_{\rm disc}/L_{\rm
      Edd})^{0.25}$. The impact of the orbital parameters (distance,
    mass and inclination angle) uncertainties on the disc luminosity
    is shown on the inset for each source (see text for
    precisions). Colour/symbol code:  GX 339-4
    (\textcolor{green}{$\triangledown$}: SWIFT
    data. \textcolor{green}{$\blacktriangledown$}: T08.
    \textcolor{green}{$\triangle$}: ASCA
    data. \textcolor{green}{\Large$\blacktriangle$\small}: M06). GRO
    J1655-40 (\textcolor{gold}{\Large$\bullet$\small}, SWIFT
    data). XTE J1817-330 (\textcolor{marroncaca}{$\square$}: SWIFT
    data. \textcolor{marroncaca}{$\blacksquare$}: S07). SWIFT
    J1753.5-0127 (\textcolor{red}{\Large$\diamond$\small}, SWIFT
    data). XTE J1118+480 (\textcolor{blue}{$\bigstar$}. open : EUVE
    data. filled : SAX data). A0620-00
    ({\bf\textcolor{purple}{\small$\hexagon$\small}}).}\label{fig-kt-ldisc-tout} 
\end{figure}

We first investigate the global relation between the disc luminosity
and its inner temperature (Fig. \ref{fig-kt-ldisc-tout}). As was
pointed out in the previous sections, a constant truncation radius
would mean a constant relationship with $T_{\rm in}\propto L_{\rm
  disc}^{0.25}$. We thus have drawn in Fig. \ref{fig-kt-ldisc-tout}
this relation, normalised to the temperatures obtained in the highest
luminosity states of GX 339-4 and XTE J1817-330. An examination of the
diagram shows that around $L_{\rm disc}\sim 10^{-2} L_{\rm Edd}$, the
points seem to slightly deviate from this relationship toward the
lower luminosities. If we then include the points at lower
luminosities and in quiescence (mostly coming from XTE J1118+480),
this deviation appears more clearly.  The fact that the points are
located below the relationship would thus indicate an increase of the
multicolour disc normalisation $K_{\rm disc}$, hence the radius
(normalised to $\rm R_{\rm g}$).

\begin{figure}
\includegraphics[width=0.5\textwidth]{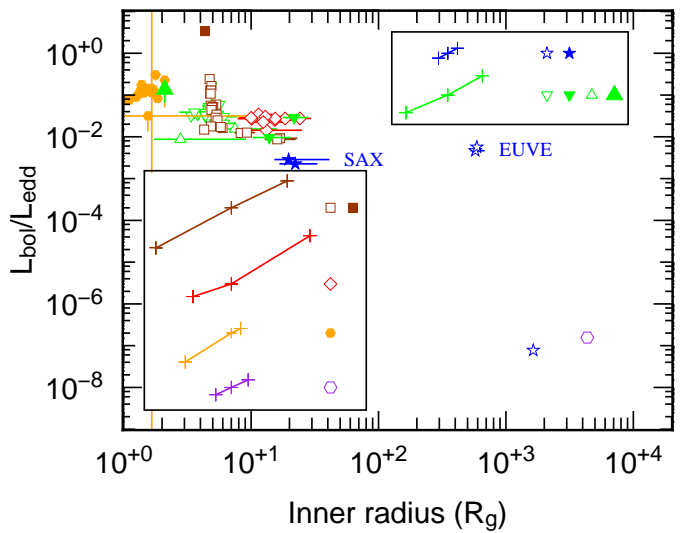}
  \caption{Inner radius versus $L_{\rm bol}/L_{\rm Edd}$. Shifts in radius and luminosities due to orbital parameters uncertainties are shown in the insets. Colour/symbol code is the same as in Fig. \ref{fig-kt-ldisc-tout}.}\label{fig-rayon-lbol-tout}
\end{figure}

As pointed out in the previous paragraph, GRO J1655-40 ($\bullet$
points) exhibits a rather different behaviour as for the same disc
(and even bolometric, see Fig. \ref{fig-rayon-lbol-tout}) luminosity,
the inner temperature is higher in the highest states. This is more
obvious when considering that their points are not located around the
dashed relation in Fig. \ref{fig-kt-ldisc-tout}. The bias coming from
the distance and the mass uncertainties could explain part of this
discrepancy. However, as the disc luminosity should be about 10 times
higher in order to fit this relation, it implies changes of a factor
three in distances or a factor 10 in masses between sources (or some
combination of the two). Another reasonable explanation could be that
the truncation radius value in the highest states of GRO J1655-40 is
closer to the horizon than in the other X-ray binaries studied,
increasing its efficiency for an equivalent accretion rate. This may
imply higher spin.

\begin{figure}
\includegraphics[width=0.5\textwidth]{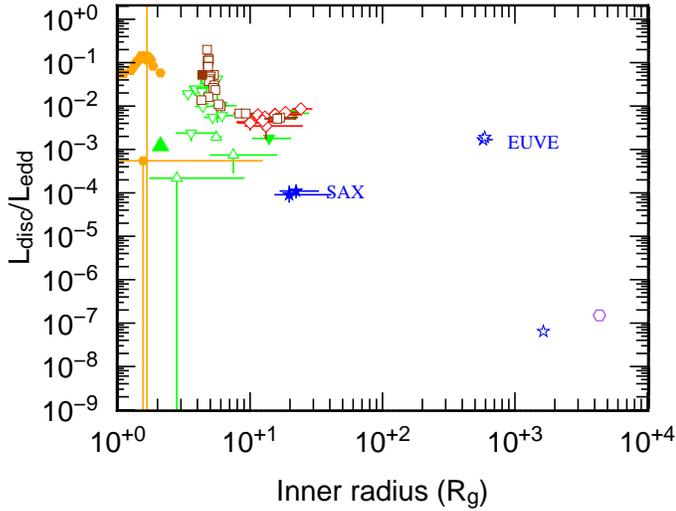}
  \caption{Evolution of the disc luminosities in function of the inferred internal radius between different sources.
}
\label{fig-rayon-ldisc-tout}
\end{figure}

\begin{figure}
\includegraphics[width=0.5\textwidth]{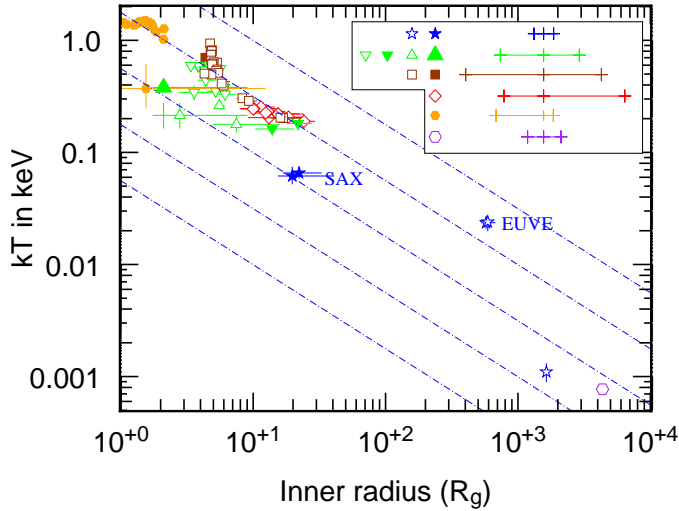}
  \caption{Inner radius versus temperature. Blue dotted-dashed lines :
    relationship obtained for a 8 $\mathrm{M_{\odot}}$ BH accreting at
    constant $\dot{m}_{\rm d}$ and an Eddington efficiency
    $\eta=0.1$. From upper right to lower left $\dot{m}_{\rm
      d}=1,\times 10^{-2},\times 10^{-4},\times 10^{-6},\times
    10^{-8}$. Colour/symbol code is the same as in
    Fig. \ref{fig-kt-ldisc-tout}.}\label{fig-rayon-kt-tout} 
\end{figure}

\begin{figure}
\includegraphics[width=0.5\textwidth]{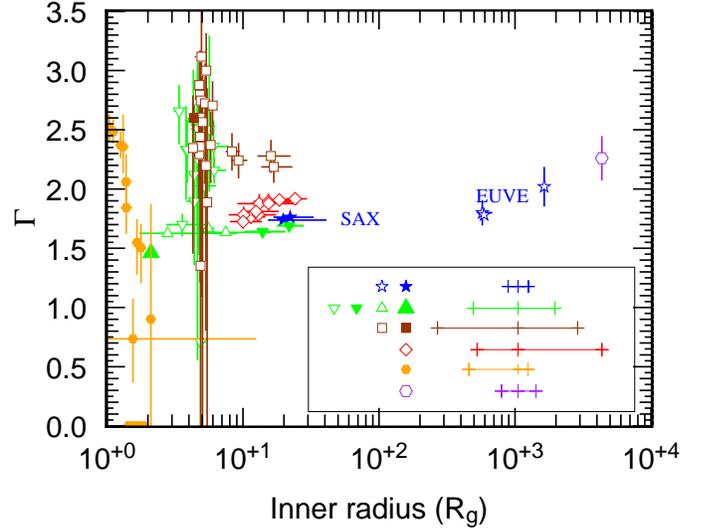}
  \caption{Photon index $\Gamma$ vs the internal radius of the
    optically thick disc. Colour/symbol code is the same as in
    Fig. \ref{fig-kt-ldisc-tout}.}\label{fig-rayon-gamma-tout} 
\end{figure}

The apparent overall increase in radius when the luminosity decreases
was then probed by the relation between the disc or bolometric
luminosity (Figs. \ref{fig-rayon-lbol-tout} and
\ref{fig-rayon-ldisc-tout}) and the inner temperature
(Fig. \ref{fig-rayon-kt-tout}) or the photon index
(Fig. \ref{fig-rayon-gamma-tout}). When examining the disc luminosity
as a function of the inner radius (Fig. \ref{fig-rayon-ldisc-tout}),
this trend seems to be confirmed with a few exceptions. Among them
are some of the GX 339-4 observations, and obviously the one
observation from XMM by M06
(\textcolor{green}{\Large$\blacktriangle$\normalsize} point).

 As pointed out by these authors, the radius value inferred by the
 thermal component is close to $2 \rm\ R_{\rm g}$, even though the disc
 luminosity is very low ($L_{\rm disc}\sim 10^{-3}L_{\rm Edd}$),
 suggesting a non recessed disc even in the low accretion rate
 states. However, considering the bolometric luminosity (see e.g
 Fig. \ref{fig-rayon-lbol-tout}), we note a different behaviour. In
 that case we have $L_{\rm bol}\sim 1.4\times 10^{-1}L_{\rm
   Edd}$. This suggests that the source was in quite a high state
 during this observation, even though being spectrally hard (as the photon
 index $\Gamma\sim 1.46$). Indeed, as shown in Fig.
 \ref{fig-rayon-lbol-tout}, and looking at whole data set, the
 evolution of $L_{\rm bol}$ vs $R_{\rm in}$ seems to be less
 discrepant and an increasing radius is consistent with $L_{\rm
   bol}\lesssim 10^{-2}L_{\rm Edd}$.

This is emphasised by the analysis of Fig. \ref{fig-rayon-kt-tout}, as
the inner radius values seem to roughly anti-correlate with the inner
temperatures when $kT_{\rm in}\lesssim 0.5\ \rm keV$. The data seems to
follow one of the dashed lines for the relation between $T_{\rm
  in}$ and $R_{\rm in}$ (Eq. \ref{eq-kt-mdot-rin}) for a
$8\mathrm{M_{\odot}}$ BH and a constant accretion rate in the disc,
$\dot{m}_{\rm d}$.

\begin{figure}
\includegraphics[width=0.5\textwidth]{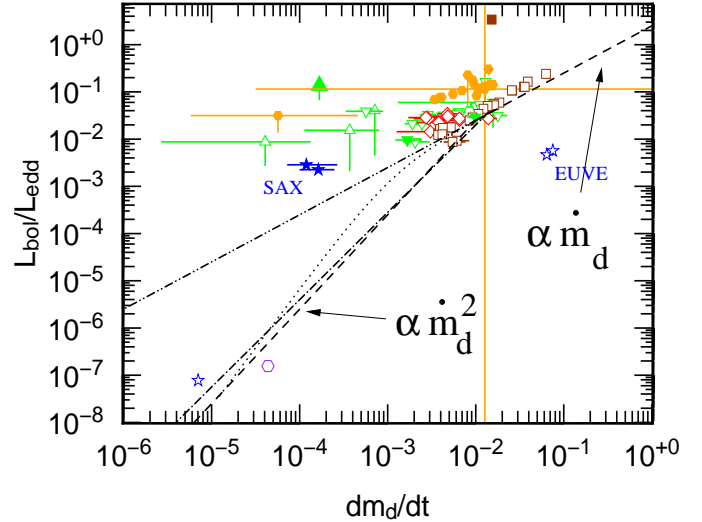}
  \caption{Bolometric luminosity as function of the accretion rate in
    the optically thick disc $\dot{m}_{\rm d}$ obtained for our
    sample, using equation \ref{eq-mdot-ldisc}. Points are subject to
    global horizontal shifts due to the actual Eddington efficiency
    value used.  The set of curves shows theoretical relations. Dashed : $L_{\rm
  bol}\sim L_{\rm disc}\propto \dot{m}_{\rm d}$ when $R_{\rm
  in}=6\rm\ R_{\rm g}$ and the efficiency at Eddington luminosity
$\eta_{\rm Edd}=0.1$. Under $\sim 10^{-2}L_{\rm Edd}$, continuing
relationship when $L_{\rm bol}\propto \dot{m}_{\rm d}^2$. Supposing
$\dot{m}_{\rm total}\sim \dot{m}_{\rm d}$ we have:  ADAF solutions
(dotted), MH02 (Dot-dot-dashed) and RC00 (Dot-dashed). The value of
$\alpha$ and $\beta$ are the same as in
Fig. \ref{fig-rayon-lbol-tout-model}.   Errors on $\dot{m}_{\rm d}$
were computed by propagating those on the temperature and the inner
radius only. 
Colour/symbol code is the same as in Fig. \ref{fig-kt-ldisc-tout}.}\label{fig-lbol-mdot}
\end{figure}

 However, by using (Eq. \ref{eq-mdot-ldisc}) and taking into account
 the different BH masses, we infer the evolution of the disc accretion
 rate $\dot{m}_{\rm d}$. As shown in Fig. \ref{fig-lbol-mdot}, the
 disc accretion rate seems to roughly scale with the bolometric
 luminosity in the highest states, i.e $L_{\rm bol}>10^{-2}L_{\rm
   Edd}$. The behaviour under this value is less clear. On the one
 hand, XTE J1817-330 seems to exhibit a beginning of rapid decay of
 the radiation efficiency ($L_{\rm bol}$ divided by a factor 2 for a
 rather constant $\dot{m}_{\rm d}$ between observation number 15 and
 17). On the other hand, considering the whole data set at those low
 luminosities, the obvious lack of disc detections and the larger
 uncertainties are insufficient to draw the profile of $L_{\rm bol}$
 as a function of $\dot{m}_{\rm d}$ with accuracy. It is however
 compatible with $L_{\rm bol}/L_{\rm Edd}\propto\dot{m}_{\rm d}^2$ by
 taking into account the sources observed in quiescence. This
 corresponds to a scenario in which a receding, but radiatively
 efficient, outer disc feeds an inner, radiatively inefficient flow.

However, the canonical BHB states do not depend on luminosity (see e.g the
classification scheme of \citealp{mcclintock03b} which is based on spectral, timing
and radio behaviour or \citealp{homan05}). Therefore we plot the
evolution of the inner radius as function of the photon index,
$\Gamma$ in Fig. \ref{fig-rayon-gamma-tout}. The evolution is then as
follows. In the softest states, radii remain quite constant, as for
example the average value obtained for XTE J1817-330 and GX 339-4 when
$\Gamma>2.4$ (14 observations) is $R_{\rm in,\ avg}=4.91 \rm\ R_{\rm
  g}$, with a standard deviation of $\sigma_{R_{\rm
  in,\ rms}}=0.59 \rm\ R_{\rm g}$. On the contrary, for lower photon
indices, the data are compatible with an increase of the inner radius
when the spectrum hardens, as long as quiescence is not
reached. Indeed, in quiescence the photon index values are usually
higher ($\Gamma\sim 2$). This effect was noted in other sources as
well (e.g. in V404 Cyg $\Gamma\sim 2.1$, see \citealp{narayan97}).

\begin{figure}
\includegraphics[width=0.5\textwidth]{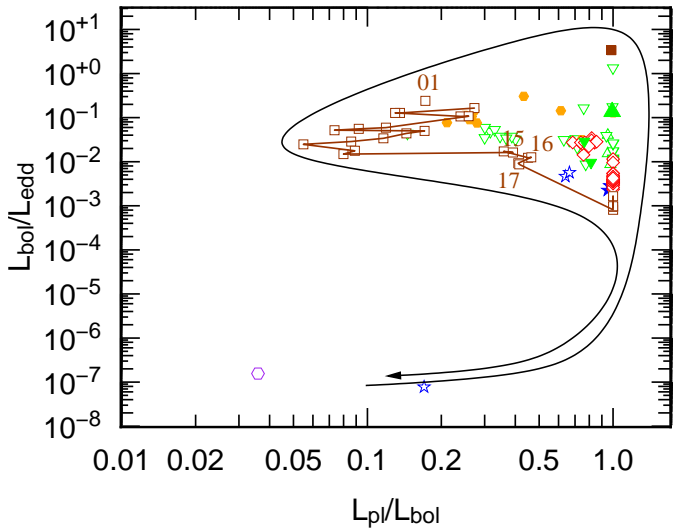}
  \caption{\rm Disc fraction luminosity diagram of the whole set of
    observations presented in this paper. The arrow represent the
    supposed path in the DFLD during an outburst, starting from
    quiescence. Colour/symbol code are the same as in
    Fig. \ref{fig-kt-ldisc-tout}.}\label{fig-dfld} 
\end{figure}

Finally, we investigate the evolution of the relative strength of each
component (i.e emission from the optically thick vs optically thin
medium) during the outburst. ``Hardness-Intensity Diagrams'' (HID) are
commonly used for such analysis because of their emission model
independence (see e.g \citealp{belloni04}).  However, they are not easily adapted
to multi-instrument and multi-object analysis. Following the method used by
\cite{koerding06}, we use the so-called ``Disc Fraction Luminosity
Diagram'' (DFLD) that aims at generalising the concept of HID: 
in HIDs harder states are located on the right of the diagram and correspond to higher powerlaw contribution. For better ressemblance, in DFLD we use on X axis the ``Power law fraction'' (PLF):
\begin{equation}
PLF=\frac{L_{\rm PL}}{L_{\rm disc}+L_{\rm PL}},
\end{equation}
which describes well the disc contribution as the ``Disc Fraction'' (DF) is then in turn:
\begin{equation}
DF=1-PLF. 
\end{equation}

The DFLD corresponding to our data set is plotted on
Fig. \ref{fig-dfld}. Of primary importance is first the pattern drawn
by XTE J1817-330 during its decline.  As the bolometric flux fades, the
optically thick component is dominating and even seems to increase its
contribution (PLF falls by a factor 2 between observation number
three and 13). Then the PLF suddenly increases and this turnover
corresponds to the observations (number 15, 16 and 17) where a
recessed disc begins to appear. However, it is not accompanied by a
drastic change in bolometric luminosity, suggesting that the inner
part of the disc has been replaced by a quite efficient corona.

We also note that the powerlaw fraction value in quiescence is equivalent
to that observed in the softest states. Therefore we suggest that the
track followed by a BH X-ray binary should be roughly equivalent to
the one plotted on Fig. \ref{fig-dfld} (black arrow). From quiescence,
the source will progressively increase its PLF as it usually enters in
outburst by its low-hard state. Then, the powerlaw fraction decreases due
to the presence of the thermal component in the former ``high-soft''
state and finally returns to the hard and quiescent states.

However, this simple pattern is susceptible to be altered whenever a
high intermediate state with a strong powerlaw component is present.
In order to confirm this trend, further monitoring is therefore still
necessary.   

\subsection{Comparison with timing semi-empirical relationship and
  accretion models}

\subsubsection{ADAF and inner disc evaporation
  models}\label{subsub-accretion-model}

Further investigations were then led to compare our results to
available accretion models. Our goal is to constrain the mechanisms
responsible for the optically thin and radiatively inefficient
flows. For that purpose, we mainly used the models proposed by
\cite[CRK04]{czerny04a} as particular attention is paid to the
evolution of inner disc radii. We compared our data to the following
three models - the classical strong ADAF solutions
(\citealt{abramowicz95} = ``A'' model), and two other models
(\citealp{meyer02} = ``B'' model or ``MH02'' and \citealp{rozanska00}
= ``C'' model or ``RC00'') where evaporation of the inner part of the
cold disc is taken into account. In these latter models, the
evaporation efficiency is driven by conduction between the corona and
the disc in presence of a magnetic field. 

We also identified the evaporation radius $\rm R_{\rm evap}$ described
in those papers with the internal radius we obtained from our fits
($\rm R_{\rm evap}=R_{\rm in}$). As modeled by CRK04, one can
then infer the dependency between $R_{\rm in}$ and $L_{\rm bol}$
simply by tuning the values of two parameters, the classical $\alpha$
viscosity parameter and the magnetisation $\beta=P_{\rm gas}/(P_{\rm
  mag}+P_{\rm gas})$ in the hot plasma.

\begin{figure}
\includegraphics[width=0.5\textwidth]{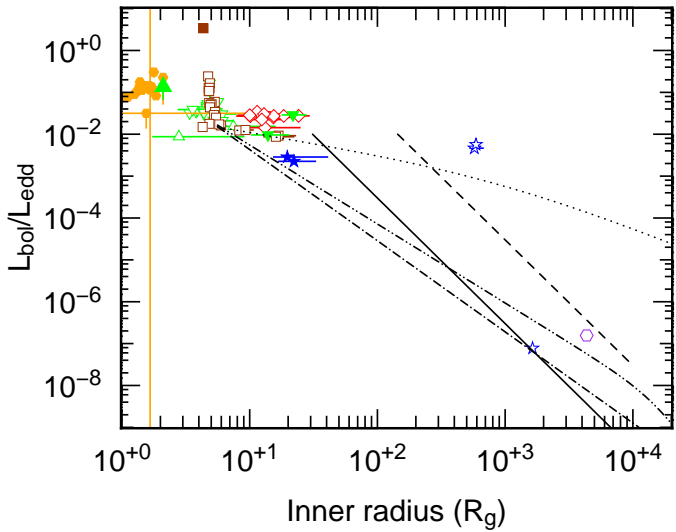}
  \caption{Same as Fig. \ref{fig-rayon-lbol-tout} but
      with models overplotted. Continuous curve refers to the
      accretion model or timing relationship used (see
      text). Parameters value used: ADAF (dotted curve)
      $\alpha=1.33\times 10^{-2}$ and $\beta=1.e-8$. MH02 (Dot-dashed
      curve): $\alpha=0.3$ and $\beta=0.11$. RC00 (Dot-dot-dashed):
      $\alpha= 3\times 10^{-5}$ and $\beta=0.11$. For the timing
      relationships, we use either $\nu_{\rm l}=\nu_{\rm
        Keplerian}(R_{\rm in})$ (dashed line) or $\nu_{\rm l}=\nu_{\rm
        sound}(R_{\rm in})$ with $h_{\rm d}/R_{\rm in}=0.1$ (solid
      line). Colour/symbol code is the same as in
      Fig. \ref{fig-kt-ldisc-tout}.}\label{fig-rayon-lbol-tout-model} 
\end{figure}

We then used the equations given in CRK04 and briefly explored the
$(\alpha,\beta)$ plane for models ``A'' and ``C'' ($\beta$ plane only
for model ``B'', as the analysis driven by \citeauthor{meyer02} was
only computed with $\alpha=0.3$). One constraint we assess is the
values of the luminosity when the inner radius seems to begin to
increase. We set this value arbitrarily to the one obtained with XTE
J1817-330 (see Tab. \ref{var-par-1817-tbl}), i.e $L_{\rm bol}\sim
0.015\ L_{\rm Edd}$ and $R_{\rm in}=5.6\rm\ R_{\rm g}$. Each model
with an example of $(\alpha,\beta)$ values that fulfil this
requirement was then overplotted on
Fig. \ref{fig-rayon-lbol-tout-model}.

 For the ADAF (``A'') model, the adopted value for the
 viscosity parameter was $\alpha=0.013$.  As shown as well
 in Fig. 2 of CRK04, the radiative efficiency depends only slightly on
 the value of the magnetisation parameter in this case. Decreasing
 $\beta$ only tends to deviate a little towards the lower luminosity
 the $L_{\rm bol}$ vs $R_{\rm in}$ curve for large values of the inner
 radius. As a result, even with such a low value as $\beta=10^{-8}$
 (i.e plasma highly magnetised) , the curvature does not allow the
 ADAF model to reproduce the very low luminosity observed in
 quiescence if the viscosity parameter $\alpha$ is the same between
 the highest and quiescent state (as shown on
 Fig. \ref{fig-rayon-lbol-tout-model}). Any attempt to fit the points
 in quiescence would require a value of $\alpha\sim 3\times 10^{-4}$,
 however the state where the radius would begin to increase would then
 occur at $L_{\rm bol}\sim\times 10^{-5}L_{\rm Edd}$. We note
 that the ADAF model fits well the beginning of the recession observed
 in XTE J1817-330. This model predicts indeed that 

\begin{equation}
\frac{R_{\rm in}}{\rm R_{\rm g}}\propto \alpha^4\dot{m}^{-2}.
\end{equation}

 Moreover, in the highest states, the accretion process still remains
 quite efficient, giving $L_{\rm bol}/L_{\rm Edd}\propto \dot{m}$ (see
 Eq. 16, 17 and 18 of CRK04 for further details). We therefore get: 

\begin{equation}
\frac{R_{\rm in}}{\rm R_{\rm g}}\propto \left(\frac{L_{\rm bol}}{L_{\rm Edd}}\right)^{-2}.
\end{equation}

 It is interesting to note that this exponent value of -2 is
 consistent with that obtained when fitting by a broken powerlaw the
 $R_{\rm in}$ vs $L_{\rm bol}$ points for XTE J1817-330 (see
 Tab. \ref{var-par-1817-tbl} : $\alpha_2=2.4$ or $1.7$). Moreover,
 when extended to larger radii, the ADAF solution seems to be in quite
 good agreement with the data obtained with EUVE for the XTE J1118+480
 outburst as well.

Nevertheless, we have shown that the ADAF solutions do not seem to
fit the $L_{\rm bol}$ vs $R_{\rm in}$ relationship in both high and
quiescent states.  On the other hand, the approximate relation giving
the accretion radiative efficiency (equations 16, 17 and 18 of CRK04)
seems to fit the data, as shown on Fig. \ref{fig-lbol-mdot} (dotted
curve) and if we assume that $\dot{m}$ scales as $\dot{m}_{\rm d}$.

Compared to the ADAF model, the models where evaporation of the inner
part of the disc is taken into account seem to be a better overall
description of the $L_{\rm bol}$ vs $R_{\rm in}$ evolution. In these
cases, the magnetisation value has far more impact. Taking
$\beta=0.11$ (i.e $P_{\rm mag}\sim 8P_{\rm gas}$) with the MH02 model
allows us to fit the overall shape of both the quiescent and the low
state observations (except XTE J1118+480 EUVE data). A similar result can be
obtained with RC00 model with a similar value of the magnetisation,
but quite a lower value of the viscosity parameter ($\alpha=3\times
10^{-5}$). The main drawback concerning such a low value of $\alpha$
can be seen on Fig. \ref{fig-lbol-mdot} (dot-dot-dashed curve). The
radiative efficiency does not drop at low $\dot{m}$ and the accretion
flow remains radiatively efficient, inconsistent with the points in
quiescence. No other $(\alpha,\beta)$ combination has been found which better
mimics the behaviour in both $L_{\rm bol}$ vs $R_{\rm in}$ and $L_{\rm
  bol}$ vs $\dot{m}$ diagrams.  This problem does not seem to occur
with MH02 model (dot-dashed curve) where the value of $\alpha=0.3$ is
higher. 

\subsubsection{Using the Variability Plane}

We subsequently tried to compare our spectral fitting results with the
expected timing behaviour. Indeed, \cite{mchardy06} found that AGN and
stellar soft state BH could be unified by taking the accretion rate
and the BH mass into account. Indeed, a fundamental plane seems to
link the typical frequency observed in the power spectra (especially
to the lower high frequency Lorentzian  $\nu_{\rm l}$), the BH mass
($M$) and the total accretion rate $\dot{m}$

\begin{equation}
\left(\frac{\nu_{\rm l}}{Hz}\right)\left(\frac{M}{10 M_{\odot}}\right)\propto \left(\frac{\dot{M}}{\dot{M}_{\rm Edd}}\right).\label{eq1}
\end{equation}

 \cite{koerding07} found
afterwhile that this relationship could be extended to hard states,
even if the normalisation is not exactly the same, neither is the 
 characteristic frequency (in soft state of BHB and AGN, it is the
 high frequency break).  If typical frequency observed is linked to a
 Keplerian motion near the internal radius of the accretion disc
 $R_{\rm in}$, we get:

\begin{equation}
\displaystyle\nu_{\rm l} \propto (M)^{1/2}(R_{\rm in})^{-3/2} \propto (M)^{-1}\left(\frac{R_{\rm in}}{\rm R_{\rm g}}\right)^{-3/2}, \label{eq2}
\end{equation}
when normalising $R_{\rm in}$ to the gravitational radius $\rm R_{\rm
  g} \propto M$. Thus replacing $\nu_{\rm l}$ in equation \ref{eq1} by
the previous value gives:

\begin{equation}
\displaystyle (M)^{-1}\left(\frac{R_{\rm in}}{\rm R_{\rm g}}\right)^{-3/2}(M) \propto \dot{m}\ \Longleftrightarrow\ \dot{m} \propto \displaystyle\left(\frac{R_{\rm in}}{\rm R_{\rm g}}\right)^{-3/2}.\label{eq3}
\end{equation}

For inefficient accretion flows that may occur at low luminosities,
one can also have:

\begin{equation}
\displaystyle \frac{L_{\rm bol}}{L_{\rm Edd}} \propto \dot{m}^2,\label{eq4}
\end{equation}
whereas for efficient ones it is proportional to $\dot{m}$.
We can thus infer a relationship between the internal radius of the
standard accretion disc (normalised to $\rm R_{\rm g}$) and the
bolometric luminosity by combining Eqs. \ref{eq3} and \ref{eq4}:

\begin{equation}
\displaystyle \frac{L_{\rm bol}}{L_{\rm Edd}} \propto \displaystyle\left(\frac{R_{\rm in}}{\rm R_{\rm g}}\right)^{-3},\label{eq5}
\end{equation}
in the inefficient case, whereas it goes as $(R_{\rm in}/\rm R_{\rm
  g})^{-3/2}$ when the flow is efficiently radiative.

The normalisation constant is actually undetermined in the
relationship linking the radius and the typical frequency
(Eq. \ref{eq2}), and is subject to variations in the other equations
used. Hence we plotted the dependencies demonstrated in equation
\ref{eq5} on Fig. \ref{fig-rayon-lbol-tout-model}, using either the
hypothesis that the frequency $\nu_{\rm l}$ used in Eq. \ref{eq1} is
exactly equal to the Keplerian frequency (Eq. \ref{eq2},  dashed
line in Fig. \ref{fig-rayon-lbol-tout-model}), or to a sound wave. In
this case, we have (see e.g. \citealp{czerny04b}, Eq. 3):

\begin{equation}
\nu_{\rm sound}=\left(\frac{h_d}{r}\right)\nu_{\rm Keplerian}.
\end{equation}

This latter relationship implies that the normalisation in
Eq. \ref{eq5} is multiplied by a factor $(h_d/r)^2$. In
Fig. \ref{fig-rayon-lbol-tout-model}, we plotted the corresponding
relationship for $(h_d/r)=0.1$ (solid line), which seems to be
more consistent with the points obtained by spectral fitting in the
$10^{-8}-10^{-2}$ luminosity range.

\section{\rm discussions}
\label{sec-discussion}

\subsection{Summary}

In this paper we focused on the question of the evolution of the
optically thick disc geometry during luminosity and state evolution
observed in black hole binaries. Estimates of the internal radii
values were obtained via modelling the thermal component in the
spectra with a multicolour disc. In this simple model the luminosity
is dominated by the internal part of the accretion disc that behaves
like a blackbody of temperature $T_{\rm in}$ and surface $4\pi R_{\rm
  in}^2$. 

 However, as the optically thick disc in a BHB is usually emitting in
 the soft X-rays or far UV, the normalisation of this component is
 strongly dependant the value of the interstellar (and internal)
 absorption along the line of sight. We studied
 four sources to determine if there were possible changes in the
 value of $N_{\rm H}$ during state transitions. According to the data
 obtained with \textit{Swift}, we demonstrated that when using a
 powerlaw for fitting the high energy component of the spectra, a
 decrease of the absorption is detected (above reasonable significance). This decreasing $N_{\rm H}$ is moreover observed when
 the source moves towards harder state.  This effect is however
 not significant when using the thermal 
comptonisation model \texttt{comptt} \citep{titarchuk94}.

Taking this trend into account, we then studied the evolution of the
disc geometry, firstly in the particular case of XTE J1817-330. We
demonstrated that this source was indeed exhibiting the beginning of a
disc recession.  This was already noted by \cite{gd08}, but without taking
into account disc irradiation, and contrary to what was stated by
R07. This turnover in the disc behaviour is statistically significant
(see Tab. \ref{var-par-1817-tbl}), even when fixing the value of the
hydrogen column density.

We then extended our study to a sample of six black hole binaries and
tried to draw conclusions on the overall trend of the disc geometry
evolution. These data, from a variety of instruments, are still consistent with a
recession of the inner disc radius when the bolometric luminosity and
the inner temperature decreases (see Fig. \ref{fig-rayon-lbol-tout}
and \ref{fig-rayon-kt-tout}) or the spectrum hardens
(Fig. \ref{fig-rayon-gamma-tout}).

\begin{figure*}

\includegraphics{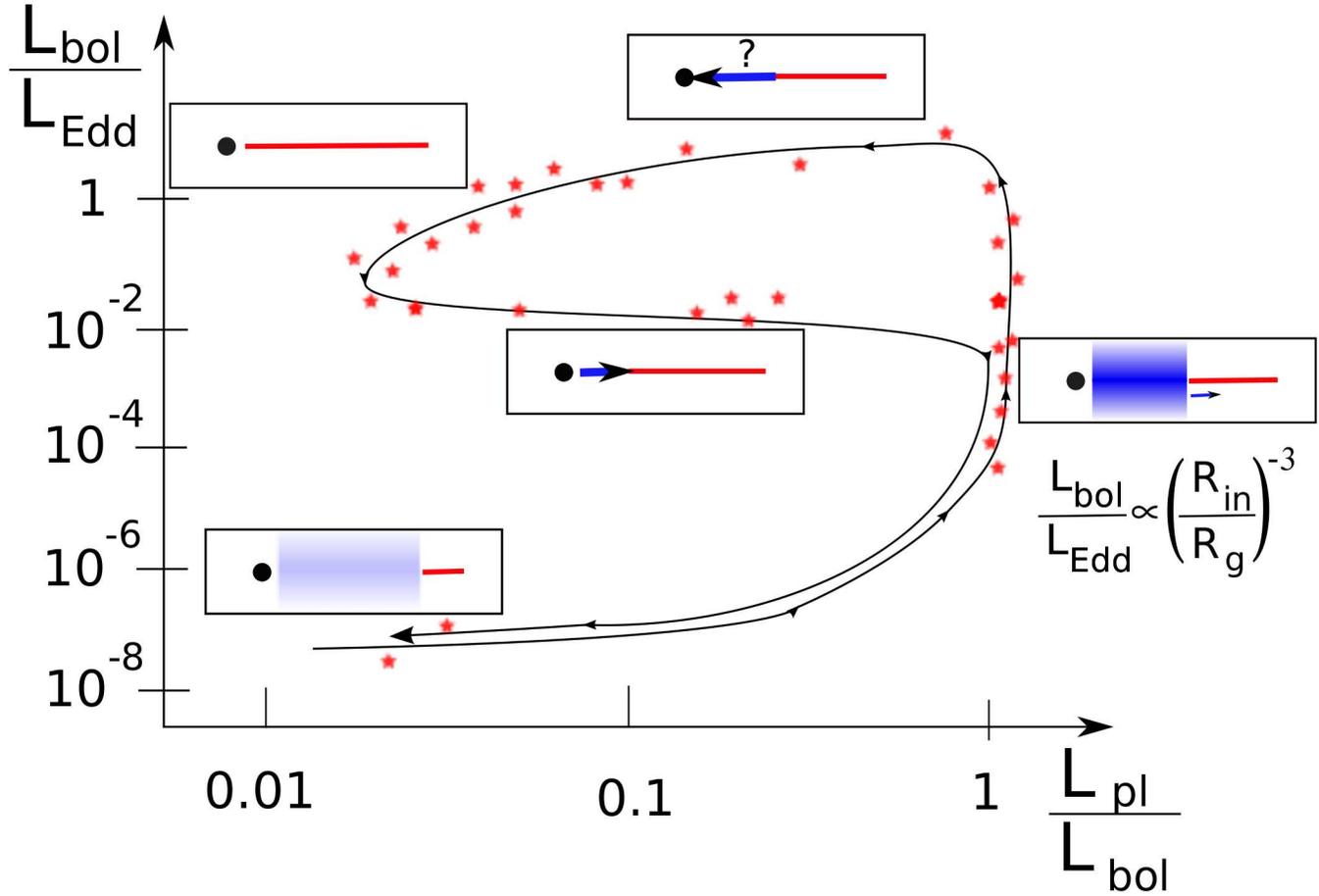}
\caption{Expected behaviour of a BHB during outburst in the DFLD. Note
  that this general shape will not be valid if the source transit by a
  very high state (or ''Steep Power Law'') with a high contribution of
  the powerlaw component.}\label{fig-cartoon} 
\end{figure*}

Moreover in order to link the quiescent and high luminosity data, we
also demonstrated that $L_{\rm bol}$ had to scale with the square of
the accretion rate flowing through the optically thick disc
$\dot{m}_{\rm d}^2$ (Fig. \ref{fig-lbol-mdot}).  Finally this
evolution is consistent with accretion models where 
 evaporation of the inner part of the disc is taken into account,
 however the strong ADAF hypothesis cannot be ruled out. The evolution
 of the disc radii values in function of the bolometric luminosity is
 also consistent with the expected behaviour derived from timing in
 the lowest accretion states (see Eq. \ref{eq5}). We therefore propose
 in Fig. \ref{fig-cartoon} a general sketch displaying the evolution
 of BHB in function of their state and accretion rate.

\subsection{Known limitations of the analysis}
We only focused our analysis on the strength of the disc component in
order to determine the disc geometry. Future investigations should
correlate these results with those obtained either by iron line
width fitting (but see hereafter the known limitations) or studies on
the timing characteristics.

 More importantly, we cannot exclude the possibility that the inner
 radii values obtained by using the simplistic multicolour disc model
 may sometimes be unphysical. Following e.g \cite{sobczak00}, a sudden
 flare could result in an apparent decrease in the inner
 disc radius value when using the simple multicolour disc to fit the
 thermal component when no such decrease takes place in reality. This
 could be due to \textit{increased spectral
   hardening and/or Compton up-scattering of soft disk photons}, thus
 depleting the thermal component, and hence decreasing the observed
 radius value. 

 However, an increase in the radius when the source fades as 
 observed in our study, seems more difficult to interpret in that
 framework. When moving towards the harder and lower states, one
 should on the contrary expect a decrease of the radius value,
 according to the previous reasons. Therefore if we had taken those
 effects into account, we would have even obtained bigger radii value
 during the transitions. This would lead to equivalent results as
 obtained by \cite{gd08}, who took into account the irradiation of the
 inner part of the disc by hard X-rays coming from the corona on the
 same XTE J1817-330 \textit{Swift} data than ours. They  concluded
 that instead of receding by a factor two or three, as we 
 obtained in this work, the inner radius would increase by a factor 6 to 8.

This irradiation model could therefore influence the relationship between
$L_{\rm bol}$ and $R_{\rm in}$, and hence, the determination of the
accretion models performed in Sect. \ref{subsub-accretion-model} via a
flattening of the actual relation. This model described in \cite{gd08}
depends mainly on one extra parameter, the ratio $R_{\rm c}=L_c/L_{\rm
  disc}$ ($L_c$ being the comptonised component luminosity), which
determines the deviation from the multicolour disc (see Fig. 6 of
\citealp{gd08}). The higher the $R_{\rm c}$ the more the deviation. As
$R_{\rm c}$ can be linked to the value of the powerlaw/disc fraction PLF
($L_c\sim L_{\rm PL}$) giving $R_{\rm c}\sim PLF/(1-PLF)$, we can
perform a rough estimate of the influence of irradiation on $R_{\rm
  in}$ in our study via the DFLD plotted in Fig. \ref{fig-dfld}. 

 For XTE J1817-330, as in observations 15, 16 and 17, $PLF\sim 0.5$ and
 therefore $R_{\rm c}=0.5$. Using Fig. 6 of \cite{gd08}, the radius
 should increase by a factor 1.25 only. This correction factor is
 negligible for points in higher states. The effect is however more
 striking for GX 339-4, SWIFT J1753.5-0127 and XTE J1118+480 in
 outburst. Indeed, both T08 data points have a $R_{\rm c}\sim 4$ and
 thus the radii obtained should then increased by a factor 3, giving
 $R_{\rm in}\sim 66$ and $42 \rm\ R_{\rm g}$. For SWIFT J1753.5-0.127,
 as $R_{\rm c}$ can reach values of 5, it would imply a correction
 factor of about 4.  The case of XTE J1118+480 data obtained with
 Beppo-SAX is even more characteristic, as the luminosity
 is completely dominated by the optically thin component and $R_{\rm
   c}\sim 20$. As the lack of coverage in Fig. 6 of \cite{gd08} does
 not allow us to infer the correction in radius, this implies, an
 increase by a factor five should be a minimum. Paradoxically, this
 could solve the discrepancy observed between radius values determined from
 Beppo-SAX and EUVE data.  The EUVE data indicate that the correction factor in
 radius should not be higher than 2 ($R_{\rm c}\sim 2$). On the
 contrary, both quiescent points might be slightly affected by
 irradiation as in that case $R_{\rm c} < 0.2$. In any case, our study
 could be improved in the future by investigating the effect of adding
 irradiation, or reflection  models to the fits (see e.g
 \citealp{reis08} for a reanalysis of M06 data with reflection models,
 and \citealp{hiemstra08} for the case of SWIFT J1753.5-0127 in the
 lowest states). 

 In addition to our extensive study, we note that an apparent disc
 recession during the decline was also noted in RXTE data at the end
 of the observations analysed by \cite{sobczak00} on XTE J1550-564
 (the inner radius value rises by a factor three when the total flux
 drops by a factor 50).

 The value of $R_{\rm in}$ obtained in quiescence can also be subject
 to systematic deviation, but for other reasons than those mentioned
 so far. As already pointed out by (MC03a), the optical/UV
 excess observed in quiescence can also be mimicked by an ADAF,
 without requiring a disc, if we use the latest version of this
 model \citep{narayan97}. The value of the transition radius $R_{\rm
   tr}$ obtained by MC03a is then $\sim 10^4 R_{\rm S}$. As those
 latter authors do, we can hence reasonably estimate that in
 quiescence $2\times 10^3 \leq R_{\rm in}/\rm R_{\rm g}\leq2\times
 10^4$. 

 Another problem arose from our study concerns the discrepancy
 observed in the outburst of XTE J1118+480 between Beppo-SAX and EUVE
 data. We already mentioned that disc irradiation could be one
 explanation, however it does not seem to be sufficient to account for
 the very large differences observed. Another possible explanation could come
 from the uncertainties in evaluating the extinction in EUVE data and
 thus the EUVE slope (see e.g Fig.3 of (MC01)), which could strongly
 affect the flux emitted by the source and hence the estimate of the
 radius. Independently to the discussions presented in (MC01) and
 (Ch03), we note that such a high radius from the EUVE data implies
 certain issues regarding the accretion
 efficiency. Fig. \ref{fig-dfld} shows that in order to account for
 the observed data, the accretion rate flowing through the disc must
 be quite high ($\dot{m}_{\rm d}\sim 0.1$). However, the low
 luminosity observed would also imply very inefficient flow, which is
 quite unusual compared to what is observed in the other BH binaries
 for such accretion rate. 

 \indent We note however that the case of XTE 1118+480 in its
   2000 outburst has recently been revisited by \cite{reis09}. These
   authors base their analysis on Chandra and RXTE data alone and
   conclude that a non recessed disc is present, attributing the soft
   excess observed in EUV and radio to the synchrotron component of
   the jet. However the handling of the absorption in this study seems
   puzzling. In the final fits, the value of the absorption seems to
   stick at its higher boundary (fixed at $1.3\times 10^{20}\ \rm
   cm^{-2}$). We note furthermore that when it is allowed to vary more
   freely, its final value is quite higher ($\sim 2.7\times
   10^{20}\ \rm cm^{-2}$).

\subsection{Consequences of a truncated/non truncated disc}

 Assuming we state that the overall trend of the optically thick disc
 is to exhibit a progressive recession of its inner radius, how can we
 interpret the recent observations of GX 339-4 (M06) and SWIFT
 J1753.5-0127 \citep{mi06b}? First, as stated by those authors
 themselves, it can be peculiar case, or perhaps observations
 performed when the source was not steady. We cannot exclude, due to
 the long duration of the observation that the state could be a
 ``mix'' of a source transiting from the low to the high state for GX
 339-4 for example. 

Regarding the GX 339-4 observations from (M06) in detail, several
 points seems interesting to investigate.  As we demonstrated
 (in Fig. \ref{fig-rayon-lbol-tout}), the bolometric luminosity is
 high ($\sim 0.1\ L_{\rm Edd}$) though with a low disc
 contribution. Hence, following \cite{gd08}, a complete treatment with
 the irradiation taken into account could give quite higher value to
 the inner radius observed as the powerlaw flux is completely
 dominating ($R_{\rm c}\sim 112$). Independently from this, the results
 obtained by the iron line fit with the \texttt{laor} model gives a
 very low inclination angle, $i=18^{\rm o}$. Thus, if the inner disc
 and the orbit of the secondary are coplanar, as the mass function is
 $f_M=M_1\sin^3(i)/(1+M_2/M_1)^2=5.8\pm0.5$ (\citealp{hynes03}, where
 $M_1$: BH mass and $M_2$: secondary mass), this would imply a black
 hole mass of $M_1\sim 200 \mathrm{M_{\odot}}$ (for
 $M_2=1\mathrm{M_{\odot}}$) or $M_1\sim 215 \mathrm{M_{\odot}}$ (for
 $M_2=10\mathrm{M_{\odot}}$). These values are quite unusual for an
 X-ray transient, and the only way to disentangle this is to suppose
 the presence of a warped disc in GX 339-4 (see e.g
 \citealp{fragile01} for possible evidences in GRO J1655-40, or
 \citealp{maccarone02} for a more extensive study).

The case of SWIFT J1753.5-0127 in \cite{mi06b} is more subtle, even if
as in GX 339-4, the powerlaw flux is dominating ($R_{\rm c} \sim 18$)
and thus the irradiation could have an effect on the evaluation of
$R_{\rm in}$. However, if the luminosity is quite low in \cite{mi06b},
it mostly relies on the values of the mass and the distance of the
source, which are currently not well known. If, we choose the values
of our study (see Tab. \ref{tab-mass-dist-angle}), we obtain $L_{\rm
  bol}/L_{\rm Edd}\sim 1.5\times 10^{-3}$. However, if $d=8.5\ \rm
kpc$, $L_{\rm bol}$ is nearly an order of magnitude higher.

We can also note that the addition of a reflection component
on the disc (such as in \citealp{reis08} or \citealp{hiemstra08})
could also explain part of the soft X-ray emission, and thus could
decrease the disc intensity and the inner radius value inferred from
the fits.

 More fundamentally, a non truncated disc in the lowest state could
 raise some issues according to \cite{dubus01} simulations. When a
 disc is present near the ISCO, instabilities in the disc seems to be
 unavoidable and a truncated disc prevents the triggering of such
 outbursts.

 Finally, if we extend this study to larger scales, some data can also
 be obtained in the gap between $10^{-3}$ and $10^{-4}$ $L_{\rm Edd}$
 thanks to X-ray-Bright Optically Normal Galaxy (XBONG) and low
 luminosity AGN (M81, NGC 4579) data. Indeed, following \cite{yuan04},
 the corresponding inner radius obtained are close to $R_{\rm in}\sim
 10^2 \rm\ R_{\rm g}$ at $L_{\rm bol}\sim 10^{-4} L_{\rm Edd}$. Of
 importance, Fig. 3 of \cite{yuan04} seems indeed to
 exhibit roughly the same $L_{\rm bol}$ vs $R_{\rm in}$ relationship
 we obtain in Fig. \ref{fig-rayon-lbol-tout-model}, the corresponding
 ``quiescent AGN'' being Sgr A$^{\rm *}$ and M87.

\section{Conclusions}
Based on several datasets obtained with instruments
sensitive at low X-ray-energies (SAX, Swift, EUVE), we studied the optically thick component in BHB spectra and its evolution during outbursts via its contribution in the soft X-ray spectra.  Any possible degeneracy with other processes such as the absorption in the line of sight was analysed. Depending on the nature of the high energy emission process, a possible variation of the hydrogen column density during outbursts is detected.\\
As a result of the good coverage by \textit{Swift}, the case of XTE J1817-330 is examined in detail. Whereas for high bolometric luminosities ($>10^{-2}\rm\ L_{Edd}$) the inferred inner radius remains fairly constant, a sudden increase of its value is significantly detected when the luminosity is low ($<10^{-2}\rm\ L_{Edd}$).
When including results coming from other sources and in quiescence as well, the  hypothesis of a beginning of disc recession under $10^{-2}-10^{-3}\rm\ L_{Edd}$ is still favoured. The observed disc recession rate with luminosity
is also consistent with X-ray timing behaviour of BHs.

\section*{Acknowledgments}
CC thanks J. Malzac, G. Dubus, T. Maccarone, S. Chaty, I. Papadakis,
P-O Petrucci, C. Done and the anonymous referee for their useful comments and
discussions. This research has been funded in part by a Philip Leverhulme Prize
awarded by the Leverhulme Foundation to RPF. RJHD acknowledges support from the Alexander von Humboldt Stiftung/Foundation. EGK acknowledges funding via a Marie Curie
Intra-European Fellowship under contract no. MEIF-CT-2006-024668.
This research was supported by the DFG cluster of excellence ‘Origin
and Structure of the Universe’ (www.universe-cluster.de).

\appendix

\addappheadtotoc

 \renewcommand{\thefigure}{\Alph{figure}}
 \setcounter{figure}{0}
\renewcommand{\thetable}{\Alph{table}}
 \setcounter{figure}{0}

\begin{table*}
\centering 
\caption{Table of the relevant fit/literature results we used in this study. The column densities values are expressed in $10 ^{22}\rm \ cm^{-2}$, $kT_{\rm in}$ in keV, $K_{\rm disc}$ in $\rm 10^3 \times (R_{\rm in,\ km}/d_{\rm 10\ kpc})^2 \cos(i)$, and the fluxes in terms of $\rm 10^{-10}\ erg.cm^{-2}.s^{-1}$. The powerlaw flux values in parenthesis were not taken into account as the photon indices values obtained are irrealistic. Absorbed powerlaw fluxes are computed between 0.05 and 200 keV where unabsorbed disc fluxes are computed on the whole electromagnetic spectrum. Values written in italics are fixed during the fitting process. 
}
\begin{tabular}{ccccccccc}
\hline
Telescope&ObsId/Ref&$N_{\rm H}$&$kT_{\rm in}$&$K_{\rm disc}$&$F_{\rm disc}$&$\Gamma$&$F_{\rm PL}$&$\chi^2/\nu$\\
\hline
\multicolumn{9}{|c}{XTE J1817-330}\\
\textit{Swift}-XRT&$00030367001$&$0.165^{+0.026}_{-0.025}$&$0.939^{+0.014}_{-0.014}$&$2.25^{+0.18}_{-0.15}$&$378.5^{+18.7}_{-15.0}$&$2.31^{+0.27}_{-0.39}$&$79.14^{+17.32}_{-67.99}$&$577.8/542$\\ 
\textit{Swift}-XRT&$00030367002$&$0.117^{+0.003}_{-0.003}$&$0.849^{+0.005}_{-0.005}$&$2.12^{+0.06}_{-0.06}$&$239.1^{+1.9}_{-1.3}$&$-$&$-$&$623.7/536$\\ 
\textit{Swift}-XRT&$00030367002$&$0.128^{+0.055}_{-0.006}$&$0.815^{+0.007}_{-0.013}$&$2.39^{+0.11}_{-0.27}$&$227.9^{+21.3}_{-17.4}$&$1.35^{+1.33}_{-1.90}$&$85.81^{+0.00}_{-85.81}$&$407.4/451$\\ 
\textit{Swift}-XRT&$00030367003$&$0.195^{+0.031}_{-0.029}$&$0.807^{+0.010}_{-0.010}$&$2.29^{+0.12}_{-0.12}$&$210.0^{+6.7}_{-6.6}$&$2.88^{+0.27}_{-0.28}$&$33.12^{+5.92}_{-9.52}$&$432.7/461$\\ 
\textit{Swift}-XRT&$00030367003$&$0.171^{+0.031}_{-0.031}$&$0.802^{+0.010}_{-0.010}$&$2.37^{+0.14}_{-0.13}$&$211.7^{+7.1}_{-6.7}$&$2.60^{+0.29}_{-0.46}$&$31.70^{+6.07}_{-13.21}$&$481.1/480$\\ 
\textit{Swift}-XRT&$00030367004$&$0.155^{+0.020}_{-0.019}$&$0.747^{+0.009}_{-0.008}$&$2.30^{+0.14}_{-0.14}$&$154.3^{+6.3}_{-6.8}$&$2.29^{+0.18}_{-0.23}$&$48.76^{+4.06}_{-6.67}$&$521.4/518$\\ 
\textit{Swift}-XRT&$00030367005$&$0.175^{+0.025}_{-0.025}$&$0.741^{+0.012}_{-0.011}$&$2.33^{+0.21}_{-0.20}$&$152.0^{+7.2}_{-8.5}$&$2.36^{+0.20}_{-0.28}$&$53.22^{+5.40}_{-6.27}$&$463.7/480$\\ 
\textit{Swift}-XRT&$00030367006$&$0.123^{+0.030}_{-0.015}$&$0.633^{+0.010}_{-0.009}$&$2.86^{+0.17}_{-0.16}$&$99.3^{+1.3}_{-0.0}$&$2.20^{+0.47}_{-1.39}$&$13.47^{+0.00}_{-13.47}$&$368.1/396$\\ 
\textit{Swift}-XRT&$00030367007$&$0.160^{+0.048}_{-0.032}$&$0.661^{+0.009}_{-0.009}$&$2.34^{+0.12}_{-0.11}$&$96.2^{+2.8}_{-2.8}$&$2.80^{+0.37}_{-0.62}$&$9.82^{+0.35}_{-0.06}$&$350.9/390$\\ 
\textit{Swift}-XRT&$00030367008$&$0.126^{+0.035}_{-0.023}$&$0.638^{+0.016}_{-0.009}$&$2.55^{+0.14}_{-0.13}$&$91.3^{+3.0}_{-2.6}$&$2.56^{+0.44}_{-2.56}$&$7.28^{+2.17}_{-7.28}$&$347.4/368$\\ 
\textit{Swift}-XRT&$00030367009$&$0.205^{+0.098}_{-0.092}$&$0.620^{+0.023}_{-0.018}$&$2.46^{+0.36}_{-0.48}$&$78.8^{+6.0}_{-6.0}$&$3.12^{+0.63}_{-2.02}$&$16.32^{+0.00}_{-6.03}$&$232.3/261$\\ 
\textit{Swift}-XRT&$00030367009$&$0.154^{+0.043}_{-0.042}$&$0.609^{+0.012}_{-0.012}$&$2.40^{+0.21}_{-0.19}$&$71.5^{+4.1}_{-3.5}$&$2.75^{+0.36}_{-0.83}$&$12.09^{+0.00}_{-1.29}$&$305.8/324$\\ 
\textit{Swift}-XRT&$00030367010$&$0.166^{+0.042}_{-0.036}$&$0.550^{+0.008}_{-0.007}$&$2.89^{+0.15}_{-0.14}$&$57.1^{+1.2}_{-1.5}$&$3.00^{+0.31}_{-0.50}$&$7.54^{+1.67}_{-2.53}$&$380.9/364$\\ 
\textit{Swift}-XRT&$00030367011$&$0.134^{+0.028}_{-0.022}$&$0.536^{+0.008}_{-0.008}$&$2.79^{+0.16}_{-0.15}$&$49.8^{+1.3}_{-1.3}$&$2.72^{+0.33}_{-0.68}$&$4.71^{+1.75}_{-4.71}$&$351.8/350$\\ 
\textit{Swift}-XRT&$00030367012$&$0.108^{+0.021}_{-0.006}$&$0.512^{+0.009}_{-0.008}$&$2.99^{+0.18}_{-0.19}$&$44.5^{+1.4}_{-1.8}$&$1.89^{+0.96}_{-2.57}$&$2.60^{+1.77}_{-2.60}$&$330.1/317$\\ 
\textit{Swift}-XRT&$00030367013$&$0.122^{+0.037}_{-0.017}$&$0.495^{+0.012}_{-0.012}$&$2.36^{+0.22}_{-0.19}$&$30.7^{+2.0}_{-1.5}$&$2.44^{+0.54}_{-1.22}$&$3.01^{+1.74}_{-3.01}$&$273.9/273$\\ 
\textit{Swift}-XRT&$00030367013$&$0.114^{+0.021}_{-0.012}$&$0.505^{+0.009}_{-0.009}$&$1.84^{+0.13}_{-0.12}$&$26.0^{+1.1}_{-1.3}$&$2.35^{+0.44}_{-0.89}$&$2.27^{+0.71}_{-2.27}$&$309.9/310$\\ 
\textit{Swift}-XRT&$00030367015$&$0.190^{+0.029}_{-0.029}$&$0.394^{+0.013}_{-0.013}$&$3.59^{+0.57}_{-0.61}$&$18.8^{+2.2}_{-2.1}$&$2.70^{+0.20}_{-0.25}$&$12.12^{+0.65}_{-1.10}$&$301.1/328$\\ 
\textit{Swift}-XRT&$00030367015$&$0.153^{+0.031}_{-0.024}$&$0.412^{+0.017}_{-0.017}$&$3.36^{+0.52}_{-0.44}$&$20.9^{+2.4}_{-2.2}$&$2.37^{+0.28}_{-0.31}$&$11.76^{+0.74}_{-0.95}$&$320.5/290$\\ 
\textit{Swift}-XRT&$00030367016$&$0.137^{+0.015}_{-0.014}$&$0.304^{+0.015}_{-0.016}$&$6.91^{+1.69}_{-1.23}$&$12.8^{+0.8}_{-1.1}$&$2.32^{+0.15}_{-0.15}$&$10.59^{+0.00}_{-0.47}$&$304.1/321$\\ 
\textit{Swift}-XRT&$00030367016$&$0.147^{+0.014}_{-0.013}$&$0.288^{+0.016}_{-0.016}$&$8.64^{+2.48}_{-1.75}$&$12.8^{+1.0}_{-1.1}$&$2.24^{+0.14}_{-0.15}$&$11.19^{+0.92}_{-0.76}$&$279.6/316$\\ 
\textit{Swift}-XRT&$00030367017$&$0.150^{+0.019}_{-0.017}$&$0.202^{+0.017}_{-0.017}$&$28.52^{+20.07}_{-10.36}$&$10.3^{+0.0}_{-0.2}$&$2.19^{+0.13}_{-0.13}$&$7.26^{+0.92}_{-0.66}$&$241.3/247$\\ 
\textit{Swift}-XRT&$00030367017$&$0.151^{+0.020}_{-0.018}$&$0.204^{+0.019}_{-0.018}$&$25.88^{+19.35}_{-9.97}$&$9.7^{+0.2}_{-0.0}$&$2.28^{+0.13}_{-0.13}$&$6.91^{+0.56}_{-0.62}$&$265.2/234$\\ 
\textit{Swift}-XRT&$00030367019$&$0.073^{+0.061}_{-0.055}$&$-$&$-$&$-$&$1.74^{+0.22}_{-0.20}$&$1.55^{+0.00}_{-0.44}$&$18.3/23$\\ 
\textit{Swift}-XRT&$00030367020$&$0.005^{+0.050}_{-0.005}$&$-$&$-$&$-$&$1.73^{+0.25}_{-0.11}$&$3.15^{+3.07}_{-1.23}$&$25.9/24$\\ 
\textit{Swift}-XRT&$00030367021$&$0.035^{+0.044}_{-0.035}$&$-$&$-$&$-$&$1.72^{+0.20}_{-0.19}$&$1.81^{+1.05}_{-0.55}$&$27.8/31$\\ 
\textit{Swift}-XRT&$00030367022$&$0.069^{+0.039}_{-0.036}$&$-$&$-$&$-$&$2.28^{+0.22}_{-0.20}$&$1.87^{+0.60}_{-0.32}$&$35.4/38$\\ 
XMM&(S07)&$0.23^{+0.01}_{-0.01}$&$0.70^{+0.01}_{-0.01}$&$1.90^{+0.05}_{-0.05}$&$99.45$&$2.6^{+0.1}_{-0.1}$&$6571.9$&$7536.25/6029$\\
\hline
\multicolumn{9}{|c}{SWIFT J1753.5-0127}\\
\textit{Swift}-XRT&$00030090001$&$0.261^{+0.036}_{-0.030}$&$0.204^{+0.033}_{-0.028}$&$81.04^{+141.77}_{-48.75}$&$30.6^{+8.0}_{-15.1}$&$1.88^{+0.07}_{-0.08}$&$96.43^{+2.39}_{-0.00}$&$385.1/397$\\ 
\textit{Swift}-XRT&$00030090003$&$0.248^{+0.022}_{-0.019}$&$0.221^{+0.022}_{-0.021}$&$110.15^{+97.31}_{-46.96}$&$56.4^{+8.0}_{-18.5}$&$1.90^{+0.06}_{-0.06}$&$166.86^{+14.16}_{-10.99}$&$541.1/542$\\ 
\textit{Swift}-XRT&$00030090006$&$0.229^{+0.015}_{-0.013}$&$0.254^{+0.018}_{-0.018}$&$61.17^{+31.68}_{-18.72}$&$55.2^{+4.3}_{-7.9}$&$1.76^{+0.05}_{-0.05}$&$249.54^{+19.93}_{-17.23}$&$729.0/643$\\ 
\textit{Swift}-XRT&$00030090007$&$0.285^{+0.013}_{-0.011}$&$0.190^{+0.009}_{-0.009}$&$272.49^{+111.52}_{-72.20}$&$76.2^{+7.5}_{-11.6}$&$1.92^{+0.03}_{-0.02}$&$166.52^{+5.71}_{-6.74}$&$913.1/717$\\ 
\textit{Swift}-XRT&$00030090009$&$0.256^{+0.017}_{-0.015}$&$0.206^{+0.015}_{-0.015}$&$159.70^{+103.54}_{-57.52}$&$62.7^{+7.5}_{-12.4}$&$1.91^{+0.04}_{-0.04}$&$178.98^{+10.32}_{-9.91}$&$702.8/653$\\ 
\textit{Swift}-XRT&$00030090010$&$0.250^{+0.026}_{-0.022}$&$0.220^{+0.026}_{-0.024}$&$111.33^{+125.42}_{-54.19}$&$56.1^{+6.1}_{-20.5}$&$1.88^{+0.06}_{-0.06}$&$185.64^{+15.59}_{-15.42}$&$590.3/540$\\ 
\textit{Swift}-XRT&$00030090012$&$0.227^{+0.017}_{-0.015}$&$0.232^{+0.019}_{-0.019}$&$78.28^{+52.24}_{-28.34}$&$49.2^{+4.8}_{-12.4}$&$1.78^{+0.05}_{-0.05}$&$220.27^{+19.37}_{-22.31}$&$743.2/642$\\ 
\textit{Swift}-XRT&$00030090013$&$0.215^{+0.020}_{-0.017}$&$0.246^{+0.028}_{-0.026}$&$47.88^{+43.45}_{-20.49}$&$38.1^{+4.2}_{-8.4}$&$1.78^{+0.06}_{-0.06}$&$201.90^{+22.41}_{-21.20}$&$569.8/554$\\ 
\textit{Swift}-XRT&$00030090015$&$0.218^{+0.017}_{-0.015}$&$0.244^{+0.022}_{-0.022}$&$46.77^{+33.47}_{-17.51}$&$36.0^{+2.2}_{-11.0}$&$1.72^{+0.05}_{-0.05}$&$216.40^{+22.16}_{-18.68}$&$738.9/644$\\ 
\textit{Swift}-XRT&$00030090016$&$0.250^{+0.023}_{-0.019}$&$0.225^{+0.023}_{-0.022}$&$72.45^{+65.66}_{-31.49}$&$40.3^{+4.0}_{-14.7}$&$1.81^{+0.06}_{-0.06}$&$158.85^{+13.94}_{-14.84}$&$527.2/557$\\ 
\textit{Swift}-XRT&$00030090019$&$0.171^{+0.008}_{-0.008}$&$-$&$-$&$-$&$1.90^{+0.03}_{-0.03}$&$101.99^{+0.00}_{-1.71}$&$524.9/456$\\ 
\textit{Swift}-XRT&$00030090020$&$0.154^{+0.011}_{-0.011}$&$-$&$-$&$-$&$1.76^{+0.04}_{-0.04}$&$84.84^{+8.83}_{-8.28}$&$371.6/334$\\ 
\textit{Swift}-XRT&$00030090021$&$0.177^{+0.008}_{-0.008}$&$-$&$-$&$-$&$1.73^{+0.03}_{-0.03}$&$48.54^{+3.03}_{-3.27}$&$536.3/481$\\ 
\textit{Swift}-XRT&$00030090022$&$0.177^{+0.008}_{-0.008}$&$-$&$-$&$-$&$1.69^{+0.03}_{-0.03}$&$42.70^{+3.09}_{-2.41}$&$565.7/525$\\ 
\textit{Swift}-XRT&$00030090023$&$0.186^{+0.006}_{-0.006}$&$-$&$-$&$-$&$1.68^{+0.02}_{-0.02}$&$47.19^{+2.20}_{-2.81}$&$609.8/600$\\ 
\textit{Swift}-XRT&$00030090024$&$0.174^{+0.007}_{-0.006}$&$-$&$-$&$-$&$1.66^{+0.02}_{-0.02}$&$39.87^{+1.91}_{-2.05}$&$595.1/592$\\ 
\textit{Swift}-XRT&$00030090026$&$0.208^{+0.009}_{-0.009}$&$-$&$-$&$-$&$1.67^{+0.03}_{-0.03}$&$27.53^{+2.69}_{-1.52}$&$554.7/522$\\ 
\textit{Swift}-XRT&$00030090027$&$0.213^{+0.020}_{-0.019}$&$-$&$-$&$-$&$1.65^{+0.06}_{-0.06}$&$29.12^{+3.62}_{-3.82}$&$227.6/254$\\ 
\textit{Swift}-XRT&$00030090028$&$0.195^{+0.004}_{-0.004}$&$-$&$-$&$-$&$1.62^{+0.01}_{-0.01}$&$30.62^{+0.73}_{-0.79}$&$902.7/759$\\ 
\textit{Swift}-XRT&$00030090030$&$0.208^{+0.009}_{-0.009}$&$-$&$-$&$-$&$1.68^{+0.02}_{-0.02}$&$25.46^{+1.89}_{-1.52}$&$565.7/558$\\ 
\textit{Swift}-XRT&$00030090031$&$0.190^{+0.010}_{-0.010}$&$-$&$-$&$-$&$1.62^{+0.03}_{-0.03}$&$28.11^{+2.52}_{-2.18}$&$536.7/466$\\ 
\textit{Swift}-XRT&$00030090032$&$0.206^{+0.017}_{-0.016}$&$-$&$-$&$-$&$1.66^{+0.05}_{-0.05}$&$21.02^{+2.85}_{-2.36}$&$306.6/306$\\ 
\textit{Swift}-XRT&$00030090033$&$0.216^{+0.016}_{-0.015}$&$-$&$-$&$-$&$1.65^{+0.04}_{-0.04}$&$23.01^{+2.51}_{-2.16}$&$286.4/326$\\ 
\textit{Swift}-XRT&$00030090034$&$0.177^{+0.019}_{-0.018}$&$-$&$-$&$-$&$1.58^{+0.05}_{-0.05}$&$29.76^{+4.81}_{-3.89}$&$272.2/284$\\ 
\hline
\end{tabular}
\label{tab-resultsa}
\end{table*}
\clearpage
\begin{table*}
\centering 
\caption{Continuing table \ref{tab-resultsa}.
}
\begin{tabular}{ccccccccc}
\hline
Telescope&ObsId/Ref&$N_{\rm H}$&$kT_{\rm in}$&$K_{\rm disc}$&$F_{\rm disc}$&$\Gamma$&$F_{\rm PL}$&$\chi^2/\nu$\\
\hline
\multicolumn{9}{|c}{SWIFT J1753.5-0127 (\textit{continuing})}\\
\textit{Swift}-XRT&$00030090037$&$0.177^{+0.007}_{-0.007}$&$-$&$-$&$-$&$1.65^{+0.02}_{-0.02}$&$28.47^{+1.43}_{-1.32}$&$610.5/582$\\ 
\textit{Swift}-XRT&$00030090038$&$0.193^{+0.010}_{-0.009}$&$-$&$-$&$-$&$1.56^{+0.03}_{-0.03}$&$35.68^{+2.54}_{-2.47}$&$509.6/514$\\ 
\textit{Swift}-XRT&$00030090039$&$0.181^{+0.010}_{-0.009}$&$-$&$-$&$-$&$1.56^{+0.03}_{-0.03}$&$34.54^{+2.17}_{-3.23}$&$511.7/514$\\ 
\textit{Swift}-XRT&$00030090040$&$0.260^{+0.236}_{-0.181}$&$-$&$-$&$-$&$1.61^{+0.40}_{-0.36}$&$31.29^{+52.70}_{-16.34}$&$12.0/9$\\ 
\textit{Swift}-XRT&$00030090041$&$0.186^{+0.009}_{-0.009}$&$-$&$-$&$-$&$1.61^{+0.03}_{-0.03}$&$33.98^{+1.74}_{-2.36}$&$538.7/513$\\ 
\textit{Swift}-XRT&$00030090042$&$0.154^{+0.009}_{-0.009}$&$-$&$-$&$-$&$1.60^{+0.03}_{-0.03}$&$40.51^{+3.16}_{-3.05}$&$522.9/495$\\ 
\textit{Swift}-XRT&$00030090043$&$0.173^{+0.009}_{-0.008}$&$-$&$-$&$-$&$1.64^{+0.03}_{-0.03}$&$31.49^{+2.38}_{-2.08}$&$593.8/512$\\ 
\textit{Swift}-XRT&$00030090044$&$0.176^{+0.008}_{-0.008}$&$-$&$-$&$-$&$1.62^{+0.02}_{-0.02}$&$33.66^{+2.15}_{-1.83}$&$650.7/553$\\ 
\textit{Swift}-XRT&$00030090045$&$0.181^{+0.008}_{-0.008}$&$-$&$-$&$-$&$1.63^{+0.03}_{-0.03}$&$35.23^{+2.52}_{-2.35}$&$573.5/526$\\ 
\textit{Swift}-XRT&$00030090050$&$0.163^{+0.009}_{-0.009}$&$-$&$-$&$-$&$1.62^{+0.03}_{-0.03}$&$37.51^{+2.77}_{-2.30}$&$486.3/486$\\ 
\hline

\multicolumn{9}{|c}{GRO J1655-40}\\
\textit{Swift}-XRT&$00030009002$&$0.574^{+0.017}_{-0.017}$&$-$&$-$&$-$&$1.62^{+0.03}_{-0.02}$&$45.54^{+2.52}_{-3.60}$&$711.8/650$\\ 
\textit{Swift}-XRT&$00030009005$&$0.662^{+0.006}_{-0.006}$&$1.482^{+0.008}_{-0.008}$&$0.64^{+0.01}_{-0.01}$&$670.7^{+2.8}_{-3.4}$&$-$&$-$&$420.6/686$\\ 
\textit{Swift}-XRT&$00030009006$&$0.658^{+0.006}_{-0.006}$&$1.486^{+0.008}_{-0.008}$&$0.62^{+0.01}_{-0.01}$&$655.6^{+2.7}_{-2.8}$&$-$&$-$&$432.1/686$\\ 
\textit{Swift}-XRT&$00030009008$&$0.767^{+0.030}_{-0.028}$&$1.510^{+0.030}_{-0.032}$&$0.25^{+0.02}_{-0.01}$&$284.0^{+9.1}_{-10.7}$&$2.62^{+0.17}_{-0.14}$&$111.42^{+12.63}_{-9.66}$&$540.1/868$\\ 
\textit{Swift}-XRT&$00030009009$&$0.805^{+0.024}_{-0.024}$&$1.392^{+0.031}_{-0.031}$&$0.36^{+0.03}_{-0.03}$&$288.7^{+8.8}_{-8.4}$&$2.48^{+0.10}_{-0.09}$&$164.22^{+22.87}_{-17.35}$&$617.2/857$\\ 
\textit{Swift}-XRT&$00030009010$&$0.787^{+0.025}_{-0.024}$&$1.439^{+0.032}_{-0.032}$&$0.31^{+0.03}_{-0.02}$&$291.5^{+7.9}_{-9.3}$&$2.51^{+0.11}_{-0.10}$&$150.93^{+24.62}_{-17.79}$&$586.0/858$\\ 
\textit{Swift}-XRT&$00030009011$&$0.741^{+0.025}_{-0.026}$&$1.365^{+0.030}_{-0.029}$&$0.46^{+0.04}_{-0.04}$&$343.0^{+8.8}_{-8.5}$&$2.37^{+0.11}_{-0.10}$&$182.59^{+26.91}_{-18.07}$&$567.0/856$\\ 
\textit{Swift}-XRT&$00030009012$&$0.761^{+0.048}_{-0.044}$&$1.401^{+0.055}_{-0.056}$&$0.50^{+0.08}_{-0.06}$&$415.5^{+22.4}_{-21.2}$&$2.36^{+0.27}_{-0.21}$&$202.51^{+74.82}_{-49.98}$&$545.1/692$\\ 
\textit{Swift}-XRT&$00030009014$&$0.593^{+0.006}_{-0.006}$&$1.461^{+0.008}_{-0.008}$&$0.59^{+0.01}_{-0.01}$&$585.6^{+2.9}_{-2.3}$&$-$&$-$&$460.2/686$\\ 
\textit{Swift}-XRT&$00030009015$&$0.749^{+0.028}_{-0.026}$&$1.364^{+0.044}_{-0.035}$&$0.80^{+0.12}_{-0.11}$&$597.3^{+87300}_{-597.3}$&$-$&$-$&$456.1/684$\\ 
\textit{Swift}-XRT&$00030009016$&$0.691^{+0.032}_{-0.032}$&$1.444^{+0.065}_{-0.058}$&$0.55^{+0.11}_{-0.08}$&$521.3^{+14.7}_{-27.3}$&$1.84^{+0.20}_{-0.21}$&$584.80^{+261.76}_{-232.97}$&$289.1/684$\\ 
\textit{Swift}-XRT&$00030009018$&$0.670^{+0.027}_{-0.025}$&$1.326^{+0.040}_{-0.031}$&$0.92^{+0.14}_{-0.13}$&$612.5^{+36.6}_{-47.5}$&$1.50^{+0.19}_{-0.29}$&$1424.60^{+469.92}_{-380.62}$&$270.9/684$\\ 
\textit{Swift}-XRT&$00030009019$&$0.721^{+0.034}_{-0.032}$&$1.461^{+0.047}_{-0.047}$&$0.55^{+0.08}_{-0.06}$&$543.8^{+147.5}_{-140.5}$&$2.06^{+0.23}_{-0.22}$&$299.57^{+128.81}_{-68.93}$&$264.7/684$\\ 
\textit{Swift}-XRT&$00030009020$&$0.621^{+0.006}_{-0.006}$&$1.520^{+0.010}_{-0.010}$&$0.65^{+0.02}_{-0.02}$&$752.7^{+3.2}_{-4.0}$&$-$&$-$&$410.9/686$\\ 
\textit{Swift}-XRT&$00030009021$&$0.712^{+0.014}_{-0.014}$&$1.452^{+0.016}_{-0.016}$&$0.78^{+0.04}_{-0.03}$&$746.1^{+5.2}_{-6.4}$&$-$&$-$&$885.2/696$\\ 
\textit{Swift}-XRT&$00030009022$&$0.660^{+0.013}_{-0.012}$&$1.400^{+0.015}_{-0.015}$&$0.85^{+0.04}_{-0.04}$&$701.8^{+5.3}_{-6.0}$&$-$&$-$&$815.3/690$\\ 
\textit{Swift}-XRT&$00030009023$&$0.677^{+0.015}_{-0.015}$&$1.437^{+0.018}_{-0.017}$&$0.70^{+0.04}_{-0.03}$&$643.2^{+6.9}_{-7.3}$&$-$&$-$&$732.8/674$\\ 
\textit{Swift}-XRT&$00030009025$&$0.681^{+0.014}_{-0.013}$&$1.191^{+0.012}_{-0.012}$&$0.99^{+0.05}_{-0.04}$&$431.0^{+3.6}_{-3.8}$&$-$&$-$&$737.8/636$\\ 
\textit{Swift}-XRT&$00030009026$&$0.716^{+0.037}_{-0.021}$&$1.023^{+0.041}_{-0.024}$&$1.28^{+0.10}_{-0.16}$&$301.8^{+27.2}_{-25.3}$&$0.90^{+0.96}_{-2.55}$&$1293.49^{+0.00}_{-1293.49}$&$641.9/556$\\ 
\textit{Swift}-XRT&$00030009027$&$1.504^{+0.536}_{-0.441}$&$-$&$-$&$-$&$0.49^{+0.23}_{-0.22}$&$593.68^{+316.52}_{-593.68}$&$53.9/47$\\ 
\textit{Swift}-XRT&$00030009028$&$1.098^{+0.688}_{-0.442}$&$0.371^{+0.232}_{-0.122}$&$0.70^{+9.67}_{-0.65}$&$2.9^{+0.0}_{-2.9}$&$0.74^{+0.33}_{-0.36}$&$237.46^{+0.00}_{-128.71}$&$52.1/46$\\ 
\textit{Swift}-XRT&$00030009029$&$0.636^{+0.105}_{-0.095}$&$-$&$-$&$-$&$1.29^{+0.13}_{-0.12}$&$30.67^{+17.35}_{-11.72}$&$109.7/100$\\ 
\textit{Swift}-XRT&$00030009031$&$0.639^{+0.740}_{-0.635}$&$-$&$-$&$-$&$0.98^{+0.54}_{-0.48}$&$2.71^{+0.00}_{-2.71}$&$10.9/9$\\ 
\hline
\multicolumn{9}{|c}{GX 339-4}\\
\textit{Swift}-XRT&$00030919001$&$0.476^{+0.075}_{-0.056}$&$0.583^{+0.017}_{-0.016}$&$3.08^{+0.45}_{-0.53}$&$76.9^{+7.5}_{-15.2}$&$2.51^{+0.43}_{-0.81}$&$20.66^{+0.00}_{-20.66}$&$332.8/332$\\ 
\textit{Swift}-XRT&$00030919001$&$0.447^{+0.084}_{-0.042}$&$0.591^{+0.020}_{-0.016}$&$3.16^{+0.37}_{-0.52}$&$83.5^{+12.8}_{-16.4}$&$2.29^{+0.56}_{-0.84}$&$23.85^{+0.00}_{-23.85}$&$299.6/356$\\ 
\textit{Swift}-XRT&$00030919002$&$0.406^{+0.034}_{-0.017}$&$0.594^{+0.018}_{-0.022}$&$3.44^{+0.53}_{-0.44}$&$92.6^{+2.6}_{-4.2}$&$(0.71^{+1.75}_{-1.86})$&$(169.44^{+0.00}_{-169.44})$&$301.6/278$\\ 
\textit{Swift}-XRT&$00030919002$&$0.437^{+0.082}_{-0.027}$&$0.601^{+0.018}_{-0.017}$&$3.10^{+0.38}_{-0.33}$&$87.7^{+9.9}_{-12.8}$&$2.26^{+0.82}_{-1.70}$&$8.67^{+0.40}_{-8.67}$&$305.8/285$\\ 
\textit{Swift}-XRT&$00030919003$&$0.457^{+0.047}_{-0.029}$&$0.596^{+0.011}_{-0.011}$&$2.10^{+0.16}_{-0.17}$&$57.2^{+5.6}_{-4.4}$&$2.33^{+0.36}_{-0.42}$&$14.23^{+0.00}_{-3.47}$&$439.5/425$\\ 
\textit{Swift}-XRT&$00030919003$&$0.461^{+0.033}_{-0.025}$&$0.591^{+0.009}_{-0.009}$&$2.18^{+0.14}_{-0.15}$&$57.5^{+4.1}_{-4.1}$&$2.33^{+0.24}_{-0.27}$&$17.70^{+0.89}_{-2.26}$&$561.4/480$\\ 
\textit{Swift}-XRT&$00030919004$&$0.504^{+0.046}_{-0.043}$&$0.595^{+0.010}_{-0.009}$&$1.66^{+0.19}_{-0.24}$&$45.0^{+4.1}_{-3.9}$&$2.66^{+0.21}_{-0.27}$&$17.46^{+0.86}_{-2.28}$&$600.5/494$\\ 
\textit{Swift}-XRT&$00030919005$&$0.434^{+0.156}_{-0.044}$&$0.559^{+0.017}_{-0.017}$&$4.62^{+0.57}_{-1.71}$&$97.7^{+0.9}_{-87.6}$&$2.50^{+0.78}_{-0.52}$&$24.39^{+5.49}_{-24.39}$&$361.4/323$\\ 
\textit{Swift}-XRT&$00030919006$&$0.421^{+0.072}_{-0.035}$&$0.547^{+0.024}_{-0.023}$&$2.77^{+0.43}_{-0.38}$&$53.7^{+6.0}_{-8.7}$&$2.12^{+0.59}_{-0.76}$&$19.69^{+6.23}_{-19.69}$&$317.6/293$\\ 
\textit{Swift}-XRT&$00030919006$&$0.461^{+0.083}_{-0.057}$&$0.543^{+0.010}_{-0.018}$&$2.69^{+0.43}_{-0.60}$&$50.4^{+8.4}_{-9.5}$&$2.57^{+0.43}_{-0.56}$&$16.98^{+1.78}_{-5.13}$&$352.3/324$\\ 
\textit{Swift}-XRT&$00030919007$&$0.411^{+0.038}_{-0.032}$&$0.441^{+0.032}_{-0.034}$&$2.79^{+0.98}_{-0.65}$&$22.9^{+3.0}_{-3.0}$&$1.90^{+0.28}_{-0.31}$&$31.38^{+8.79}_{-11.96}$&$281.8/291$\\ 
\textit{Swift}-XRT&$00030919007$&$0.457^{+0.042}_{-0.039}$&$0.379^{+0.037}_{-0.043}$&$5.35^{+3.03}_{-1.60}$&$24.0^{+4.0}_{-4.7}$&$2.36^{+0.21}_{-0.23}$&$27.34^{+2.30}_{-2.99}$&$314.1/305$\\ 
\textit{Swift}-XRT&$00030919008$&$0.427^{+0.024}_{-0.022}$&$0.351^{+0.024}_{-0.025}$&$3.92^{+1.58}_{-1.02}$&$12.9^{+1.3}_{-1.4}$&$2.03^{+0.12}_{-0.12}$&$21.45^{+2.95}_{-3.27}$&$402.7/415$\\ 
\textit{Swift}-XRT&$00030919008$&$0.441^{+0.022}_{-0.021}$&$0.331^{+0.023}_{-0.024}$&$5.44^{+2.31}_{-1.41}$&$14.0^{+1.0}_{-1.3}$&$2.16^{+0.10}_{-0.10}$&$20.43^{+1.79}_{-1.28}$&$574.9/496$\\ 
\textit{Swift}-XRT&$00030919009$&$0.406^{+0.021}_{-0.021}$&$-$&$-$&$-$&$2.36^{+0.06}_{-0.06}$&$22.99^{+1.77}_{-1.42}$&$256.3/241$\\ 
\textit{Swift}-XRT&$00030919010$&$0.353^{+0.029}_{-0.026}$&$0.343^{+0.049}_{-0.046}$&$1.86^{+1.93}_{-0.84}$&$5.6^{+0.7}_{-1.8}$&$1.70^{+0.08}_{-0.09}$&$49.65^{+8.54}_{-5.85}$&$487.6/464$\\ 
\textit{Swift}-XRT&$00030919011$&$0.344^{+0.051}_{-0.049}$&$0.626^{+0.144}_{-0.100}$&$0.08^{+0.08}_{-0.04}$&$2.5^{+0.2}_{-0.8}$&$(0.07^{+0.33}_{-0.48})$&$(1811.88^{+0.00}_{-1811.88})$&$158.7/165$\\ 
\textit{Swift}-XRT&$00030919011$&$0.431^{+0.074}_{-0.062}$&$0.459^{+0.093}_{-0.079}$&$0.22^{+0.33}_{-0.12}$&$2.1^{+0.3}_{-0.8}$&$(0.66^{+0.24}_{-0.28})$&$(236.73^{+268.48}_{-236.73})$&$168.6/162$\\ 
\hline
\end{tabular}
\label{tab-resultsb}
\end{table*}
\clearpage
\begin{table*}
\centering 
\caption{Continuing table \ref{tab-resultsb}.
}
\begin{tabular}{ccccccccc}
\hline
Telescope&ObsId/Ref&$N_{\rm H}$&$kT_{\rm in}$&$K_{\rm disc}$&$F_{\rm disc}$&$\Gamma$&$F_{\rm PL}$&$\chi^2/\nu$\\
\hline
\multicolumn{9}{|c}{GX 339-4 (\textit{continuing})}\\
\textit{Swift}-XRT&$00030919012$&$0.353^{+0.024}_{-0.023}$&$-$&$-$&$-$&$1.84^{+0.06}_{-0.06}$&$16.23^{+2.75}_{-1.91}$&$242.8/233$\\ 
\textit{Swift}-XRT&$00030919013$&$0.305^{+0.042}_{-0.038}$&$-$&$-$&$-$&$1.65^{+0.09}_{-0.09}$&$15.26^{+3.95}_{-3.48}$&$91.2/100$\\ 
\textit{Swift}-XRT&$00030919013$&$0.209^{+0.148}_{-0.135}$&$-$&$-$&$-$&$1.46^{+0.34}_{-0.31}$&$23.79^{+41.85}_{-11.42}$&$23.9/16$\\ 
\textit{Swift}-XRT&$00030919013$&$0.307^{+0.050}_{-0.046}$&$-$&$-$&$-$&$1.64^{+0.12}_{-0.11}$&$15.09^{+4.16}_{-3.49}$&$69.3/77$\\ 
\textit{Swift}-XRT&$00030943002$&$0.229^{+0.028}_{-0.026}$&$-$&$-$&$-$&$1.37^{+0.06}_{-0.06}$&$14.31^{+3.24}_{-2.32}$&$192.8/189$\\ 
\textit{Swift}-XRT&$00030943002$&$0.273^{+0.030}_{-0.028}$&$-$&$-$&$-$&$1.42^{+0.07}_{-0.06}$&$12.32^{+4.06}_{-1.77}$&$160.8/184$\\ 
\textit{Swift}-XRT&$00030953003$&$0.255^{+0.039}_{-0.035}$&$-$&$-$&$-$&$1.34^{+0.08}_{-0.08}$&$16.09^{+4.30}_{-3.27}$&$116.7/132$\\ 
\textit{Swift}-XRT&$00030953007$&$0.305^{+0.043}_{-0.039}$&$-$&$-$&$-$&$1.39^{+0.08}_{-0.08}$&$34.66^{+9.43}_{-5.74}$&$99.5/132$\\ 
\textit{Swift}-XRT&$00030953012$&$0.295^{+0.031}_{-0.029}$&$-$&$-$&$-$&$1.48^{+0.06}_{-0.06}$&$35.93^{+5.49}_{-5.35}$&$185.9/184$\\ 
\textit{Swift}+XTE&Spec. 1 (T08)&$0.850^{+0.070}_{-0.070}$&$0.178^{+0.007}_{-0.007}$&$76^{+39}_{-26}$&$27.3$&$1.69^{+0.01}_{-0.01}$&$27.00$&$532.0/211$\\ 
\textit{Swift}+XTE&Spec. 2 (T08)&$0.900^{+0.110}_{-0.110}$&$0.157^{+0.011}_{-0.011}$&$35^{+30}_{-18}$&$4.6$&$1.64^{+0.02}_{-0.02}$&$9.20$&$280.1/211$\\ 
ASCA G+S&42010010& \textit{0.43} &$0.214^{+0.060}_{-0.049}$&$1.13^{+4.99}_{-0.84}$&$0.51^{+0.12}_{-0.51}$&$1.63^{+0.02}_{-0.03}$&$10.03^{+0.03}_{-0.06}$&$1030.70/954$\\
ASCA G+S&42010000& \textit{0.43} &$0.171^{+0.042}_{-0.033}$&$7.76^{+31.84}_{-5.88}$&$1.44^{+0.00}_{-1.44}$&$1.63^{+0.01}_{-0.02}$&$17.51^{+0.61}_{-0.47}$&$1201.70/1174$\\
ASCA G+S&43001000& \textit{0.43} &$0.279^{+0.009}_{-0.009}$&$2.68^{+0.48}_{-0.38}$&$3.43^{+0.20}_{-0.02}$&$1.67^{+0.01}_{-0.01}$&$46.33^{+0.84}_{-0.87}$&$2506.09/1469$\\
XMM+XTE&M06&$0.37^{+0.05}_{-0.05}$&$0.38^{+0.01}_{-0.01}$&$0.64^{+0.08}_{-0.08}$&$2.9$&$1.46^{+0.01}_{-0.01}$&$156.26$&$3899.2/2256$\\
\hline
\multicolumn{9}{|c}{XTE J1118+480}\\
SAX L+M+P&21173001&$0.013^{+0.002}_{-0.002}$&$0.062^{+0.012}_{-0.009}$&$780.65^{+1653.72}_{-33.54}$&$2.42^{+1.36}_{-0.77}$&$1.748^{+0.007}_{-0.007}$&$37.13^{+0.52}_{-0.69}$&$213.65/189$\\
SAX L+M+P&211730012&$0.011^{+0.001}_{-0.001}$&$0.066^{+0.007}_{-0.006}$&$992.35^{+931.64}_{-471.18}$&$2.96^{+0.40}_{-0.86}$&$1.762^{+0.006}_{-0.007}$&$28.73^{+0.37}_{-0.58}$&$237.99/189$\\
EUVE+XTE&(MC01)&\textit{0.013}&$0.024^{+0.002}_{-0.004}$&$7.{\rm e}5$&$51.91$&$1.782^{+0.005}_{-0.005}$&$41.7$&$-$\\
Idem+CXC+SAX&(Ch03)&$0.011^{+0.002}_{-0.002}$&$0.024^{+0.002}_{-0.03}$&$6.6^{+2.2}_{-0.9}{\rm e}5$&$44.85$&$1.8^{+0.1}_{-0.1}$&$40.$&$-$\\
CXC+HST&(MC03a)&\textit{0.012}&$1.1^{+0.2}_{-0.2}{\rm e}{-3}$&$5.4{\rm e}6$&$1.72{\rm e}{-3}$&$2.02^{+0.16}_{-0.16}$&$1.8{\rm e}{-4}$&$-$\\
\hline
\multicolumn{9}{|c}{A0620-00}\\
CXC+HST&(MC03a)&\textit{0.019}&$7.7{\rm e}{-4}$&$2.8{\rm e}8$&$2.3{\rm e}{-2}$&$2.26^{+0.18}_{-0.18}$&$4.3{\rm e}{-4}$&$-$\\
\hline
\label{tab-resultsc}
\end{tabular}
\end{table*}

\label{lastpage}

\end{document}